\documentclass[preprint,12pt]{elsarticle}
\usepackage[margin=1in]{geometry}
\usepackage{float}

\usepackage[table]{xcolor}
\usepackage{amsmath,amsfonts}
\usepackage{algorithmic}
\usepackage{algorithm}
\usepackage{array}
\usepackage{textcomp}
\usepackage{stfloats}
\usepackage{url}
\usepackage{verbatim}
\usepackage{tabularx}
\usepackage{graphicx}
\usepackage{natbib}
\usepackage{subcaption}
\usepackage{booktabs}
\usepackage{multirow}
\usepackage{threeparttable}
\usepackage[table]{xcolor}
\usepackage{rotating}
\usepackage{float}
\usepackage[hidelinks]{hyperref}
\usepackage{bm}
\usepackage{amssymb}  
\usepackage{pifont}   
\newcommand{\cmark}{\ding{51}}     
\newcommand{\xmark}{\ding{55}}     
\newcommand{\hmark}{\ding{109}}    
\long\def\symbolfootnote[#1]#2{\begingroup%
\def\thefootnote{\fnsymbol{footnote}}\footnote[#1]{#2}\endgroup}

\journal{Reliability Engineering \& System Safety (RESS)}

\begin{document}


\title{Grounding Time-Series Foundation Models in Digital Twin Topology for Predictive Maintenance}

\author{
        Sizhe Ma$^{*}$,
        Katherine A. Flanigan$^{*,\dagger}$,
        Mario Berg\'es$^{*,\ddagger}$
        \\
        \small{$^{*}$Department of Civil \& Environmental Engineering, Carnegie Mellon University}\\
        \small{5000 Forbes Ave., Pittsburgh, PA USA 15213}\\
        \small{\texttt{\{sizhem, kflaniga, mberges\}@andrew.cmu.edu}}
}

\begin{abstract}
Digital twins increasingly support downstream analytical tasks that depend on time-series data, motivating interest in time-series foundation models (TSFMs) as scalable backbones. However, TSFMs are primarily pretrained for temporal continuation and often underperform on unseen tasks such as regression, and systematic empirical comparisons against state-of-the-art dedicated models in digital twin contexts remain limited. This paper makes three contributions. First, we benchmark five well-known TSFMs with frozen backbones on remaining useful life (RUL) prediction using the C-MAPSS dataset, finding that multivariate architectures substantially outperform univariate ones, particularly under varying operating conditions. This raises a deeper question: when cross-channel dependencies can be modeled through pretrained weights, target-task adaptation, and digital twin-derived representations, how much does each contribute, and are they complementary? Second, we propose a topology-informed fusion approach in which topological constraints, derived from the asset structure the digital twin stores among its information models, explicitly shape cross-attention, so that fused representations respect the physical system's local connectivity rather than relying on unconstrained all-to-all interactions. Third, we conduct an ablation study across C-MAPSS subsets of varying operational complexity that isolates the three sources and their interactions. The sources prove complementary rather than redundant, and topology-constrained attention outperforms unconstrained fusion, though by a small margin, enabling a frozen TSFM informed by digital twin representations to remain competitive or in some cases exceed state-of-the-art performance on this regression task.
\end{abstract}


\maketitle
\symbolfootnote[0]{\hspace*{-7mm} \textsuperscript{$\dagger$}Corresponding author\\
\textsuperscript{$\ddagger$}Mario Berg\'es holds concurrent appointments at Carnegie Mellon University (CMU) and as an Amazon Scholar. This manuscript describes work at CMU and is not associated with Amazon.}


\section*{Highlights}
\begin{itemize}
    \item Multivariate pretraining is a key driver of TSFM robustness across conditions
    \item Three cross-channel dependency sources at digital twin downstream prove complementary
    \item Topology-constrained attention lowers NRMSE 2.7--4.2\% over unconstrained fusion
    \item Topology-grounded frozen TSFM matches end-to-end models, narrows multi-condition gap
\end{itemize}

\section{Introduction} \label{sec:1}
\begin{figure*}[t!]
    \includegraphics[width=\linewidth]{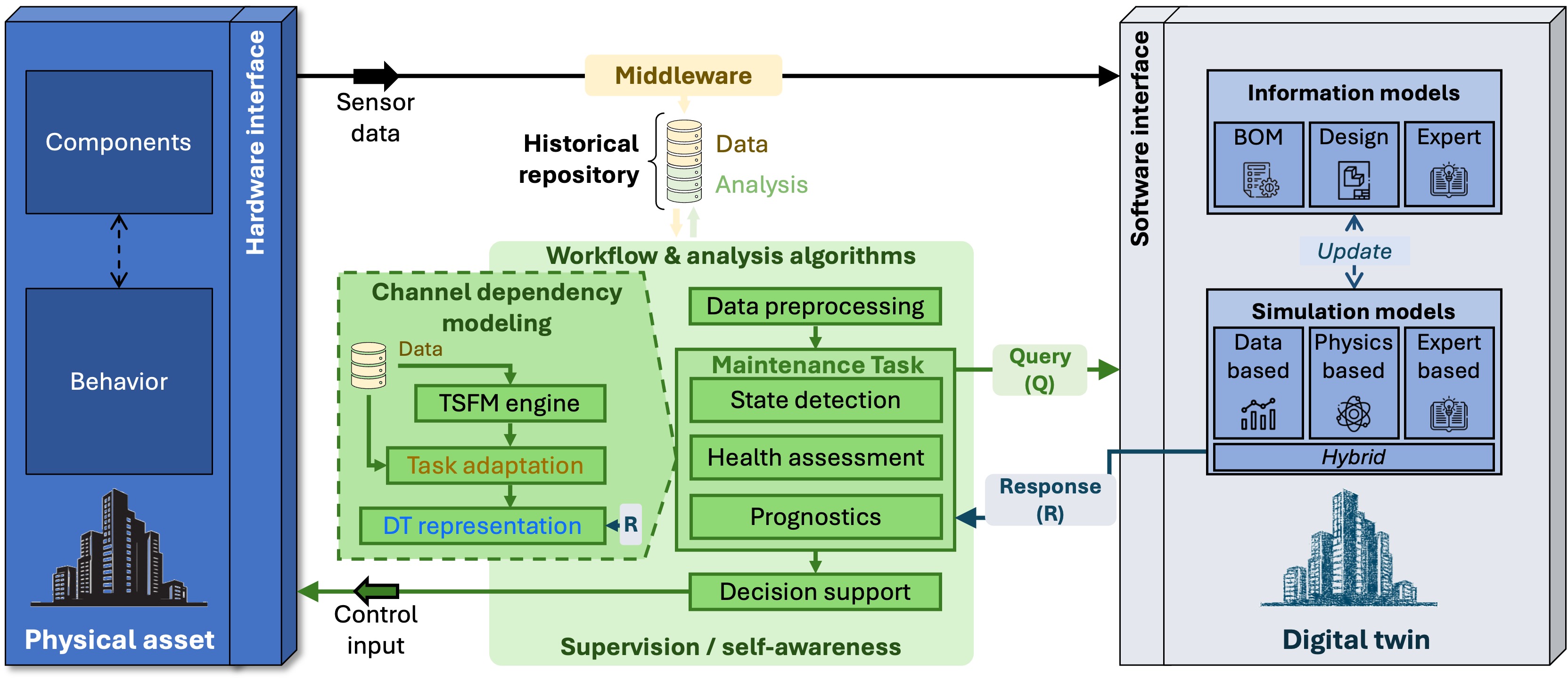}
    \centering
    \caption{Digital twin framework for standardized development.}
    \label{fig:1}
\end{figure*}

Predictive maintenance has become a cornerstone of modern industrial asset management, offering the potential to proactively reduce unplanned downtime and extend asset lifespan through data-driven decision making \cite{zonta2020maintenance}. As industrial assets grow in scale, expanding both in sensing instrumentation and operational complexity, automating predictive maintenance becomes a necessity, as only automation can sustain the uptime, cost efficiency, and operational efficiency that manual approaches cannot deliver at scale \cite{ma2025autonomy}. Sustaining this at scale, across assets operating under diverse and shifting operational conditions, demands a platform that can continuously deliver up-to-date predictive insights on demand, rather than relying on periodic analyses triggered manually. Digital twins have emerged as a leading solution for enabling this automation, providing high-fidelity virtual replicas of physical assets that continuously mirror their operational state \cite{kunzer_digital_2022}. The value of a digital twin, however, extends well beyond mirroring. One key difference that separates digital twins from related solutions, such as cyber-physical systems (CPS), is what their downstream layers are designed to do with the information they collect. CPS-based solutions have been applied to predictive maintenance tasks, but their reliance on a fixed, predefined set of analytical objectives makes them inherently rigid: adapting to a new analytical objective typically requires replacing task-specific models, reconfiguring data pipelines, or modifying system architecture \cite{baheti2011cyber}. By contrast, a predictive maintenance digital twin environment continuously integrates rich, asset-specific information that can be queried to support a broad and evolving set of analytical objectives, such as state detection, fault diagnosis, and prognostic health estimation, as shown in Figure~\ref{fig:1}, and to do so reliably across the diverse operating conditions under which the physical asset operates \cite{ma2025framework}. 

In practice, a large proportion of the analytical objectives at the downstream of a predictive maintenance digital twin are not simple information retrieval tasks but multivariate time-series tasks carried out by models within the workflow and analysis algorithms module \footnote{The workflow and analysis algorithms module governs how incoming data is processed and transformed into actionable insights \cite{gratius2024digital}.} that draw on information from both the digital twin and historical data repositories, as shown in Figure~\ref{fig:1}. These models must be capable of reasoning over correlated sensor streams evolving through time \cite{cattaneo2019twin}. A compounding challenge is that even the same task may require dedicated models\footnote{A dedicated model is a specialized algorithm designed to perform exactly one specific task, rather than a broad, multi-purpose one. This concept is also referred to as a ``special-purpose algorithm'' in computer science and ``narrow artificial intelligence'' in machine learning.} under different operating conditions, as the statistical characteristics of sensor data shift across regimes. The result is a library of such models that each must be designed, trained, and updated separately, a challenge that grows directly with the scale of sensing instrumentation and operational complexity that motivates predictive maintenance automation in the first place \cite{kapteyn2021probabilistic}.

Foundation models have emerged as a promising path out of this fragmentation, offering general-purpose backbones capable of adapting to diverse downstream tasks~\cite{touvron2023llama,radford2021learning}. At their core, FMs are large-scale neural networks pretrained on broad, diverse datasets to learn transferable representations, which can then be efficiently adapted to a wide variety of downstream tasks with minimal task-specific retraining. Originating in natural language processing (NLP) and computer vision, this paradigm is now reshaping industrial systems through time-series foundation models (TSFMs)~\cite{liang2024foundation}. For the substantial class of downstream tasks that depend on temporal data, TSFMs offer a scalable alternative to the library of dedicated models otherwise required to serve various operating conditions in the maintenance pipeline: by leveraging massive pre-training on diverse time-series datasets, a single frozen backbone, one whose pretrained weights remain fixed while only lightweight task-specific heads are trained, in principle, serve the evolving analytical platform of a predictive maintenance digital twin~\cite{liang2024foundation}. However, the current landscape of TSFMs remains heavily biased toward forecasting objectives, with most architectures pretrained for temporal continuation tasks such as next-token prediction or signal reconstruction \cite{liang2024foundation}. The tasks demanded by predictive maintenance often require the model to infer the current or future health state of an asset rather than extrapolate its future signal trajectory, a fundamentally different form of reasoning \cite{mandelli2024model,lee2023rul,gao2024domain}. Thus, whether this potential is realized on the tasks predictive maintenance digital twins actually require is a separate question.

One of the critical factors driving model performance in predictive maintenance across operating conditions is cross-channel dependencies \cite{li2021attention}. In the context of a predictive maintenance digital twin, cross-channel dependencies refer to the structured relationships among sensor channels that primarily arise from the physical, logical, or electrical coupling of components within an asset \cite{aivaliotis2019twin}. Because these relationships are rooted in the underlying mechanical, thermodynamic, or electrical interactions of the system, they remain comparatively stable as operating conditions vary---unlike raw sensor measurements, whose amplitude and distribution can shift substantially across regimes \cite{hoon2006shm}. This condition-invariance makes cross-channel dependencies particularly discriminative for tasks that need to generalize across operating conditions, such as distinguishing degradation- and condition-induced signal change \cite{zhang2025cross}. TSFMs with multivariate architectures encode a form of cross-channel dependency in their pretrained weights, learned through attention mechanisms over large-scale, cross-domain corpora \cite{ye2025time}. This raises a natural question: \textit{are the cross-channel dependencies captured in a TSFM's pretrained weights, shaped by heterogeneous data spanning unrelated domains, sufficient for reliable performance on downstream predictive maintenance tasks that demand this kind of cross-condition robustness?}

To investigate, we benchmark five TSFMs spanning a range of architectures and cross-channel modeling capabilities on C-MAPSS, a well-established multivariate benchmark providing run-to-failure trajectories for turbofan engines across both single- and multi-condition operating regimes~\cite{frederick2007user}. The downstream task is remaining useful life (RUL) estimation, a regression task representative of prognostic objectives commonly encountered at the downstream of predictive maintenance digital twins~\cite{hu2023remaining}. Using frozen pretrained backbones to generate embeddings within a unified evaluation pipeline, we conduct controlled experiments across dataset subsets of varying operational complexity. 

The results reveal that TSFMs capable of modeling cross-channel dependencies substantially outperform those that process channels independently, particularly on subsets spanning multiple operating conditions. Yet even the strongest TSFMs remain well behind state-of-the-art dedicated models trained specifically for this task, with the gap widening on multi-condition subsets. This pattern suggests that while pretrained cross-channel structure provides meaningful information, pretraining weights alone are insufficient.  However, pretrained weights are not the only source of cross-channel dependency available when a TSFM is deployed at the downstream of a predictive maintenance digital twin---two additional sources can contribute,  as illustrated in the Channel Dependency Modeling block of the workflow and analysis algorithms module in Figure~\ref{fig:1}.

Beyond pretrained weights, which serve as the first source that models cross-channel dependecy, the second source is the cross-channel dependecy learned directly from the target task and encoded in the trainable weights introduced during lightweight adaptation for downstream tasks \cite{liang2024foundation}. Even when the TSFM backbone remains frozen, adaptation mechanisms, such as learned channel-mixing layer or, in some architectures, trainable prompts, are optimized on the downstream training data and can discover how sensor channels relate to one another for the task objective at hand \cite{gao2024time}. Where the first source draws on heterogeneous corpora spanning unrelated domains, the second source is inherently specific to the deployment context: the model learns cross-channel dependency from exactly the asset and task it will be applied to. The limitation is the mirror image of the first source's: because these weights are optimized entirely on the available training samples, the cross-channel relationships they capture are bounded by whatever statistical patterns happen to appear in the training data, with no guarantee that those patterns reflect the system's physical reality or generalize beyond the operating conditions observed.

Since this paper targets predictive maintenance within a digital twin environment, a third source is available: cross-channel dependency specified directly by the digital twin, drawn from asset-specific knowledge the digital twin already encodes about the physical system \cite{thelen2023comprehensive}. This knowledge takes many forms, including physics-based equations, semantic ontologies, maintenance logs, and topological structures, all grounded in domain expertise and physical principles rather than statistical patterns in data \cite{ma2024state,ma2024state2}. The contrast with the first two sources is fundamental: where pretrained and target-task structure are \textit{learned} from observations, digital twin-derived structure is \textit{given} by the system's documented design, specifying directly which sensors monitor which components and how those components physically interact. The strength of this source is that its relationships are interpretable and asset-specific by construction; the limitation is that its fidelity is bounded by how faithfully the digital twin's structured knowledge representations, such as the information models and simulation models shown in Figure~\ref{fig:1}, model the actual asset.

The three sources thus occupy three distinct positions in the design space. Whether they capture redundant or complementary information has direct implications for how digital twin-enabled analytical systems should be designed. If the sources are laregely redundant, practitioners can rely on whiever is most convenient.  If, however, each source contributes cross-channel dependency that the others cannot provide, then neglecting any one of them sacrifices information no other source can supply. This is the question the paper sets out to answer: \textit{do these three sources contribute complementary cross-channel relationships, or do they capture redundant information?}

To investigate this question, we focus on topological graph structures as the form of digital twin knowledge under study: a representation of which sensors monitor which components and how those components are physically coupled, maintained among the information models the digital twin holds of the asset, as shown in Figure~\ref{fig:1}. Such structures are widely used within digital twins for representing the interdependencies among components in industrial assets, for example, component connectivity in manufacturing systems \cite{hawkridge2019tying}, hierarchical component in aerospace assets \cite{wang2025bayesian}, and network topology in energy infrastructure \cite{kalinin2022protection}.  Building on the controlled experimental setup from the initial benchmark on the C-MAPSS dataset, a widely adopted, well-studied benchmark in the predictive maintenance research community \cite{ramasso2014performance}, we build upon the common cross-attention-based fusion approach, and propose a topology-informed fusion in which topological constraints derived from the asset's hierarchical graph explicitly shape the attention mechanism. The fused representations respect the physical system's local connectivity rather than relying on all-to-all\footnote{All-to-all means every sensor embedding attends to every other with equal potential} interactions. The same fusion then doubles as the experimental vehicle for an ablation study that systematically isolates the three sources of cross-channel dependency and their interactions. Our results reveal that the three sources are complementary rather than redundant, each contributing cross-channel dependencies that the others do not capture. As a result, a frozen TSFM informed by digital twin-derived representation matches or exceeds state-of-the-art dedicated models on this regression task. Topology-constrained attention also outperforms unconstrained attention consistently, indicating that the physical connectivity encoded in the digital twin provides a meaningful inductive bias. Taken together, these findings provide empirical evidence for the previously theoretical position that grounding TSFMs in digital twin representations enhances their generalizability across operating conditions \cite{shen2025position}.

The primary contributions of this paper are as follows:
\begin{itemize}
    \item An empirical benchmark of five TSFMs with frozen pretrained backbones on RUL prediction, providing the first systematic comparison of how cross-channel dependencies encoded in pretrained weights, across a range of TSFM architecture, contribute to performance on a downstream predictive maintenance task demanding cross-condition robustness.

    \item A topology-informed fusion approach in which topology-derived structural knowledge, taken from the asset’s hierarchical graph as the digital twin stores it among its information models, explicitly shapes attention mechanisms, so that sensor-level representations incorporate physically grounded cross-channel dependencies rather than unconstrained all-to-all interactions.

    \item An ablation study that isolates three sources of cross-channel dependency, pretrained weights, target-task adaptation, and digital twin-derived topology, across C-MAPSS subsets of varying operational complexity, demonstrating that the three sources are complementary and that topology-constrained attention consistently outperforms unconstrained fusion.
\end{itemize}

The remainder of this paper is organized as follows. Section~\ref{sec:2} surveys the landscape of TSFMs and their cross-channel modeling capabilities, existing approaches to modeling cross-channel dependencies in predictive maintenance, and how digital twin representations have been used for knowledge grounding---establishing the gap that motivates connecting these threads. Section~\ref{sec:3} defines the case study and describes the C-MAPSS benchmark dataset, establishing the experimental foundation shared across both empirical studies. Section~\ref{sec:4} presents Case Study~I, benchmarking five off-the-shelf TSFMs on RUL prediction and analyzing how cross-channel dependency modeling in pretrained weights drives performance differences across architectures and operating conditions. Section~\ref{sec:5} introduces the proposed topology-informed fusion approach, detailing the encoding of asset topology via hierarchical message passing and its integration with TSFM embeddings via topology-constrained cross-attention. Section~\ref{sec:6} presents Case Study~II, an ablation study that systematically isolates the three sources of cross-channel dependency and benchmarks the proposed approach against state-of-the-art baselines. Section~\ref{sec:7} summarizes the findings and outlines directions for future work.

\section{Related Work} \label{sec:2}
\subsection{Landscape of Time-Series Foundation Models} \label{sec:2.1}
The emergence of TSFMs highlights a paradigm shift in analyzing temporal data, moving away from dedicated models toward universal representations learned from large-scale pretraining \cite{ye2025time}. Analogous to FMs in NLP and CV, TSFMs are pre-trained on vast repositories of diverse time-series data, sometimes exceeding billions of time points, to encode temporal dynamics, such as seasonality, trends, and local smoothness \cite{miller2024survey}. For digital twins, this paradigm promises a robust downstream capability where a single backbone can theoretically support diverse queries, across various operating conditions, related to time-series data, ranging from anomaly detection to prognostic health management, via lightweight fine-tuning or prompt learning rather than maintaining a fragmented library of specialized models \cite{shen2025position}. To illustrate this landscape, a few well-known architectures with pretrained weight available that characterize the current state-of-the-art are shown in Table~\ref{tab:1}. Goswami et al. proposed MOMENT, which adapts the masked modeling approach of BERT \cite{devlin2019bert} to the time-series domain \cite{goswami2024moment}. By employing a T5-based encoder to reconstruct randomly masked patches of the signal, MOMENT learns bidirectional dependencies and structural coherence without requiring explicit labels. Woo et al. introduced Moirai, a masked encoder-based universal forecasting Transformer pre-trained on the Large-scale Open Time Series Archive (LOTSA) containing over 27 billion observations \cite{woo2024unified}. It utilizes a multi-patch size projection strategy and Any-variate Attention to handle heterogeneous frequencies and arbitrary numbers of covariates in a zero-shot manner. Rasul et al. developed Lag-Llama, which focuses strictly on probabilistic forecasting by employing a decoder-only Llama architecture \cite{rasul2023lagllama}. This model treats lagged values as covariates, explicitly encoding strong inductive biases for periodicity and seasonality directly into the tokenization process. Ansari et al. presented Chronos \footnote{A subsequent version, Chronos~2 \cite{ansari2025chronos2}, has since been released with an expanded pretraining corpus and architectural refinements; as it became available after the experiments of this work was completed, the original checkpoint is used throughout, and Chronos~2 may yield stronger results and require further investigation.}, which reframes time series as a language modeling problem by quantizing continuous values into a fixed vocabulary of discrete bins \cite{ansari2024chronos}. This discretization allows the model to leverage off-the-shelf large language model (LLM) architectures to predict the next token via categorical cross-entropy, effectively bypassing specific time-series architectural modifications. Finally, Das et al. proposed TimesFM, a decoder-only architecture utilizing a patch-based tokenization strategy \cite{das2024decoder}. Trained on over 100 billion real and synthetic time points, TimesFM optimizes efficient context processing to scale long-horizon forecasting capabilities. \ref{app:1} summarizes the pretraining corpus each of these models reports, together with how far that corpus goes toward familiarity with the data used in this study.

\begin{table}[t]
\footnotesize
\centering
\caption{Comparison of TSFMs: MOMENT, Moirai, Lag-Llama, Chronos, and TimesFM.}
\label{tab:1}
\begin{tabular}{@{}lllll@{}}
\toprule
\textbf{Model} & \textbf{Architecture} & \textbf{Tokenization} & \textbf{Variate Handling} & \textbf{Tasks} \\ \midrule
\textbf{MOMENT}    & Encoder-only    & Patch-wise    & Univariate    & F, I, D, C \\
\textbf{Moirai}    & Encoder-only    & Patch-wise    & Multivariate  & F          \\
\textbf{Lag-Llama} & Decoder-only    & Point-wise    & Univariate    & Prob. F    \\
\textbf{Chronos}   & Encoder-Decoder & Quantization  & Univariate    & Prob. F    \\
\textbf{TimesFM}   & Decoder-only    & Patch-wise    & Univariate    & F          \\ \bottomrule
\end{tabular}%
\vspace{0.3cm}

\small \raggedright \textit{Notes}: Point Forecast (F), Probabilistic Forecast (Prob. F), Imputation (I), Anomaly Detection (D), Classification (C).
\end{table}

Despite these differences in architecture, the current landscape of TSFMs remains heavily biased toward forecasting~\cite{liang2024foundation}. The vast majority of these models are pre-trained on objectives that prioritize temporal continuation, such as next-token prediction or signal reconstruction. While models like MOMENT incorporate auxiliary tasks during pre-training, it is non-trivial to exhaustively cover the full variety of objectives encountered at the downstream of a predictive maintenance digital twin. This limitation has formal theoretical grounding~\cite{hanneke2020nofreelunch}: inductive biases shaped by temporal continuation objectives often penalize performance on tasks requiring fundamentally different reasoning, such as classification or regression~\cite{yuan2023model}.

On the other hand, the tasks demanded by predictive maintenance are predominantly health assessment tasks, including anomaly detection \cite{mandelli2024model}, fault classification \cite{lee2023rul}, and prognostic regression \cite{gao2024domain}, which require the model to infer the current or future health state of an asset rather than extrapolate its future signal trajectory. Among the information that proves critical for multivariate health assessment across operating conditions, cross-channel dependencies stand out as a particularly important source \cite{zhang2025cross}. As a physical asset operates under varying conditions, raw sensor measurements shift in distribution, making individual channel readings unreliable indicators of health state in isolation \cite{hoon2006shm}. Cross-channel dependencies, the structured cross-sensor relationships rooted in the physical coupling of components, are more stable across these shifts, because they reflect the underlying mechanical, thermodynamic, or electrical interactions of the system \cite{fassi2024physics}. Degradation originating in one component propagates towards nearby components, manifesting as correlated deviations across multiple sensors, and it is precisely these cross-channel patterns that carry the discriminative information needed to assess asset health reliably across conditions. TSFMs with multivariate architectures encode a form of cross-channel dependency in their pretrained weights, learned through attention mechanisms over large-scale, cross-domain corpora spanning domains unrelated to industrial asset health. Whether these pretrained cross-channel dependencies are sufficient for reliable performance on downstream predictive maintenance health assessment tasks, where cross-condition robustness depends on physically grounded cross-channel structure rather than statistical co-variation patterns from heterogeneous data, remains an open empirical question that existing work has not systematically addressed~\cite{ye2025time, xu2025context}. Indeed, it is perhaps surprising that TSFMs capture any meaningful cross-channel structure at all, given the breadth and heterogeneity of physical phenomena their pretraining corpora span.

\subsection{Knowledge Fusion with Time-Series Model in Digital Twins} \label{sec:2.2}
In fact, pretrained weight from TSFM is not the only source that models cross-channel dependencies at the downstream of predictive maintenance digital twin. Here, the fidelity when executing time-series tasks also relies heavily on the form and depth of the knowledge digital twin encodes about the physical system \cite{ma2024state2}. Without this asset-specific knowledge, the downstream analytical models operating within the digital twin lack the grounding needed to reason about future system states or generalize to unseen operating conditions, thus unable to finish the loop between physical assets and digital twins from a digital-to-physical perspective \cite{willcox2023foundational}. In the meantime, due to the wide applicability of digital twin across domains, the specific form of this knowledge varies significantly, ranging from single specialized models to complex collections of information representations depending on the asset's operational needs. For predictive maintenance in built environment, Eneyew et al. proposed the BIM-IoTDI framework, which leverages the rigid, object-oriented schemas of building information modeling (BIM) as the primary knowledge source \cite{eneyew2022smart}. By treating the BIM file as a static repository of geometric and semantic truth, they developed a middleware known as query mediation layer that dynamically links these spatial definitions with real-time IoT streams, enabling the digital twin to answer complex location-based queries without duplicating data. For aviation PMx, Kapteyn et al. introduced a probabilistic graphical model foundation that treats the governing laws of physics, specifically partial differential equations, as the core knowledge representation \cite{kapteyn2021probabilistic}. By compiling these physics models into a library of reduced-order models, their framework allows the digital twin to perform Bayesian inference on structural health, effectively learning the physical state of a UAV wing by matching sensor data against pre-computed physical manifolds. For predictive maintenance in manufacturing, Aivaliotis et al. introduced a framework that treats physics-based simulation models of machine components---encoding the kinematic, dynamic, and degradation behavior of each subsystem---as the core knowledge representation~\cite{aivaliotis2019twin}. By continuously tuning these models against real sensor and controller data to maintain their fidelity to the actual machine, the framework enables the digital twin to assess component health and estimate RUL directly from simulation outcomes, without requiring historical failure data. 

Within the predictive maintenance digital twin paradigm, graph structures have emerged as the common kinds knowledge across applications for modeling the cross-channel dependencies between sensors within industrial systems because they naturally encode topological relationships that unstructured data streams cannot capture \cite{schroeder2021digital}. This structural form serves as a kind of knowledge, transforming raw time-series data into representations that reflect the physical connectivity within and across asset. For instance, in application related to energy, Yang et al. utilized rigorous defect knowledge graphs to extract representations on power grid equipment and manufacturers, enabling logic-based reasoning for fault diagnosis \cite{yang2022defect}. In manufacturing, Banerjee et al. employed a semantic query mechanism to extract representations from large-scale production line data, such as the tree-like topology of the shop floor, to enhance manufacturing process management with reasoning capabilities \cite{banerjee2017generating}. In aviation, the graph-based probabilistic graphical model of Kapteyn et al. described above similarly encodes the coupled relationships between structural states, sensor observations, and control inputs, enabling Bayesian inference over health states as the asset sustains in-flight damage \cite{kapteyn2021probabilistic}.

While graph structures and other digital twin representations has the potential to model cross-channel dependencies, the question of how to effectively fuse these sources with TSFM remains open. The baseline for this integration is established by physics-informed machine learning (PIML), which typically employs constraints to guide the learning process \cite{raissi2019learning}. Zhang et al. proposed a physics-guided convolutional neural network (PhyCNN) for seismic response modeling, which integrates the equation of motion governing structural dynamics directly into the loss function to constrain the learning process \cite{zhang2020physics}. By enforcing these physical rules, the model accurately predicts structural displacements using only limited acceleration measurements, significantly outperforming standard data-driven CNNs in data-scarce regimes. Qin et al. proposed an inverse physics-informed neural network that embeds a bearing fault dynamic model, encoding mechanical parameters such as stiffness and damping ratio, directly into the network architecture as the core knowledge representation \cite{qin2024neural}. By treating the spectral discrepancy between simulated and measured vibration signals as the training objective, the framework identifies asset-specific dynamic parameters and generates high-quality fault samples across working conditions, improving cross-condition bearing fault diagnosis under imbalanced data. Kim et al. introduced a PINN-based prognostics framework that encodes low-fidelity physical knowledge, such as the monotonic and nonlinear increasing trend of degradation, as constraints during neural network training in the extrapolation region \cite{kim2022data}. By guiding the network to satisfy physically meaningful behavioral bounds even in the absence of run-to-fail data, the framework substantially reduces prediction uncertainty and yields more robust RUL estimates than purely data-driven approaches. These methodologies collectively demonstrate that informing architectures with physical laws or dynamic constraints significantly enhances robustness and generalization, particularly in regimes where data is sparse or the system is governed by well-defined dynamics \cite{djeumous2022neural}. Existing work has also extended the use of these physical laws or dynamic constraints to ground TSFMs. Sayghe et al. introduced GridFM, which utilizes physics-informed adapters to enforce power flow equations on a frozen foundation model \cite{sayghe2026gridfm}. Although this successfully grounds TSFM in the physics of the power grid, it necessitates explicit knowledge of the governing partial differential equations, a requirement that is sometimes unmet in the digital twin context, where the exact physics of every component may be unknown or too complex to formulate analytically \cite{tao2017shop}. Consequently, the pre-training of these physics-informed TSFMs poses a critical dependency on explicit knowledge of the governing equations underlying the time-series. 

Meanwhile, explicit equations are not the only approach available to inform time-series models; instead of implementing a physics-guided loss function, another common approach is to inform time-series' embeddings with knowledge via a physics-guided architecture, effectively treating knowledge as an auxiliary feature or representation source \cite{willard2022machine}. This motivates researchers to leverage architecture with attention mechanism to dynamically inform time-series embeddings with structured extracted knowledge \cite{vaswani2017attention}. Zhang et al. proposed a macro-micro attention mechanism for bearing RUL prediction that fuses a physics-guided degradation feature set, derived from the asset's digital twin and selected via an improved self-organizing feature mapping method, with temporal representations extracted from vibration signals, allowing the model to amplify physically meaningful degradation patterns across multiple scales \cite{zhang2022rolling}. Ye et al. proposed a digital-twin-assisted improved Transformer for fault prediction of automatic transverse robots in flexible packaging workshops, where a virtual replica of the robot, synchronised with the physical asset via real-time sensor streams, provides structural context for the prediction model \cite{ye2025digital}. The model integrates digital twin state outputs with multidimensional sensor time-series through an encoder-decoder cross-attention architecture, in which decoder queries attend globally to all encoder representations. Despite these successes, attention-based fusion often carries a fundamental limitation. The attention mechanism relies on all-to-all interactions, which inherently contradicts the sparse, local nature of physical interactions where components primarily influence their immediate neighbors \cite{yuan2025survey}; This architectural differences often lead to ``shortcut learning'' where the model captures rules that perform well on standard benchmarks instead of the true underlying relationship \cite{geirhos2020shortcut}. This limitation becomes particularly relevant when fusing digital twin representations with TSFMs, as the topological structure of the physical system provides explicit locality constraints that unconstrained attention mechanisms ignored. 

\subsection{Modeling Cross-Channel Dependencies in Predictive Maintenance} \label{sec:2.3}
For TSFMs with multivariate modeling capability, the pretrained weights already encode cross-channel structure: the attention mechanisms internalize statistical patterns about how sensor channels---equivalently, variables or time-series streams---tend to co-vary across the pretraining corpora \cite{woo2024unified}. The central assumption is that these corpora contain cross-channel structures relevant to the downstream task \cite{miller2024survey}. However, because the corpora span unrelated domains, the encoded weights aggregate co-variation patterns from energy grids, transportation networks, and retail sales indiscriminately, making it non-trivial to attribute learned patterns to specific physical origins. Moreover, unlike text-based models where attention can be probed for interpretable constructs such as syntax or named entities, time-series attention lacks semantic grounding, and no established probing techniques exist to characterize what cross-channel structure a TSFM has internalized \cite{derose2021attention}. Consequently, while this source enables zero- or few-shot deployment, the cross-channel structure it captures can neither be traced to specific origins nor verified against the actual system.

When labeled training data exists for the downstream task, adaptation mechanisms introduce a deployment-specific source of cross-channel structure. Even with a frozen backbone, learned channel-mixing layers or, in some architectures, trainable prompts are optimized directly on the downstream data, discovering how sensor channels relate to one another for the specific task objective \cite{benechehab2025adapts}. The central assumption is that the training data statistics are representative of the deployment conditions, a direct contrast to pretrained weights shaped by heterogeneous, multi-domain corpora. However, because these weights are optimized entirely on available training samples, the cross-channel dependency they capture is bounded by the statistical patterns the training data exhibits \cite{han2024capacity}, with no guarantee that these patterns reflect the physical reality of the system or generalize beyond observed operating conditions.

When the downstream task sits within a digital twin environment, cross-channel structure can be derived directly from asset-specific knowledge encoded in the digital twin, whose quality depends on the fidelity with which it models the actual asset \cite{austin2020architecting}. Unlike the previous two sources, which infer cross-channel dependencies from data, this source directly specifies physical interdependencies, such as which sensors monitor which components and how components physically interact, providing relationships that are interpretable and asset-specific by construction \cite{ma2024state2}. The assumption is that the digital twin faithfully represents the physical system, ensuring cross-channel relationships are grounded in physical reality rather than inferred from statistical patterns. However, this source is inherently tied to the specific asset the digital twin models and requires digital twin infrastructure to exist, which may not be available in all deployment contexts. Whether and how these three sources interact when simultaneously available is the central empirical question addressed in later sections.

\section{Case Study Definition and Dataset Description} \label{sec:3}
Investigating the contribution of cross-channel dependency modeling to TSFM performance on predictive maintenance tasks, and how digital twin representations can augment it, requires a benchmark that simultaneously provides multivariate sensor streams with known physical interdependencies, failure data captured under various operating conditions, and an established baseline of state-of-the-art performance for reference. C-MAPSS satisfies all of these requirements: it offers correlated multivariate sensor data from a physically structured turbofan engine whose component topology is well-documented \cite{frederick2007user}, spans subsets of systematically varying operational complexity that enable controlled analysis of cross-condition robustness \cite{vollert2021cmapss}, and has been extensively benchmarked in the predictive maintenance literature \cite{ramasso2014performance}. These properties make it a natural choice for both the off-the-shelf TSFM benchmarking in Case Study~I (Section~\ref{sec:4}) and the digital twin-informed fusion ablation in Case Study~II (Section~\ref{sec:6}).

C-MAPSS is a widely adopted benchmark in the predictive maintenance literature, providing run-to-failure trajectories for a fleet of turbofan engines under varying operational and fault conditions. It is an established dataset that enables direct comparison with prior state-of-the-art methods. Its four subsets ($FD001$--$FD004$) systematically vary in operational complexity and fault variability, enabling analysis of model robustness across distinct industrial scenarios. Notably, despite this ubiquity, we find no evidence of C-MAPSS itself being included in the pretraining corpus of any of the five TSFMs evaluated in this work, as discussed in \ref{app:1}. 

\begin{figure*}[t!]
    \includegraphics[width=\linewidth]{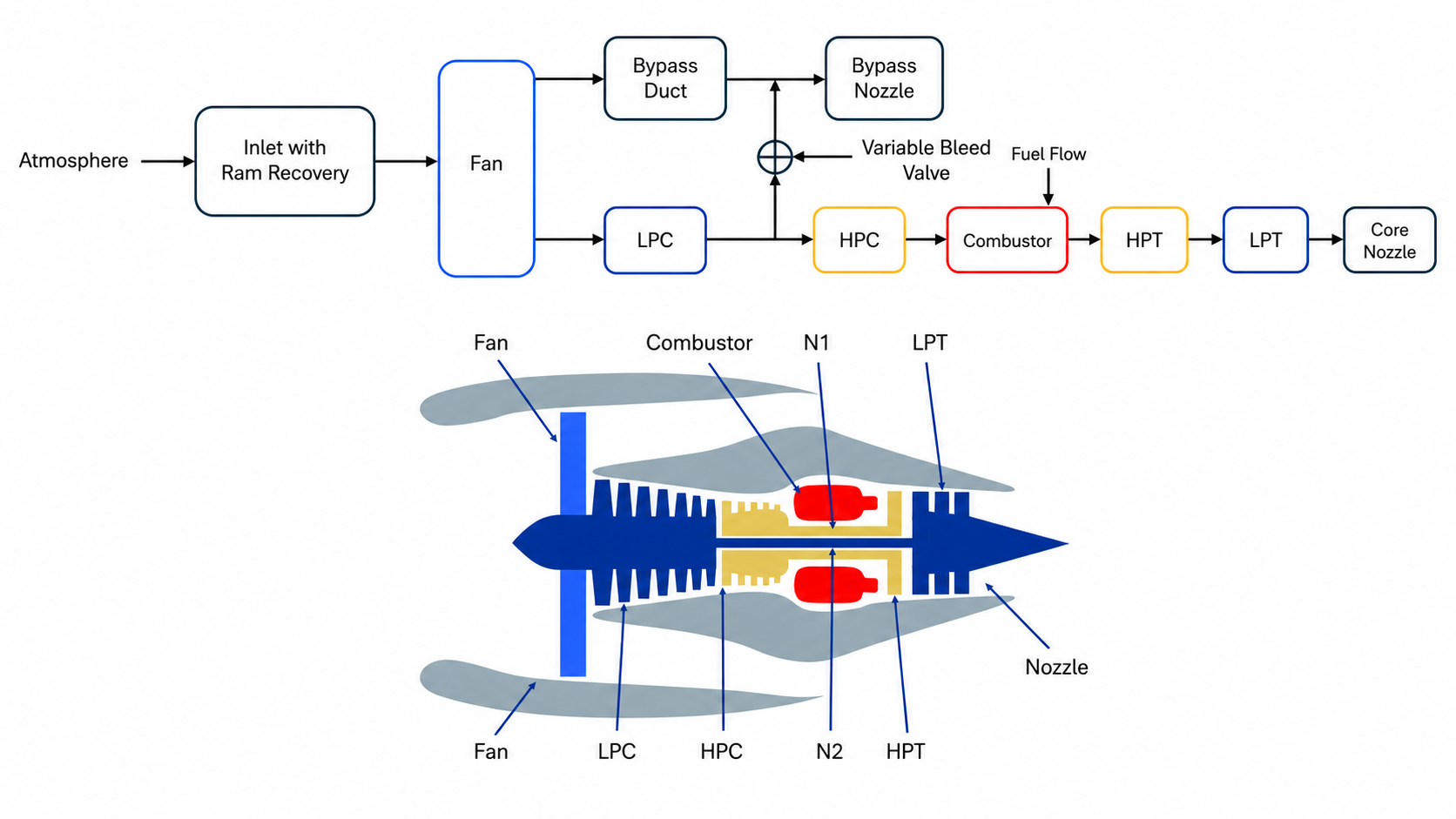}
    \centering
    \caption{System logic block diagram \cite{baptista2021map} and corresponding turbofan cross-sectional schematic for large-scale commercial turbofan engine (90,000 lb thrust class) \cite{frederick2007user}.}
    \label{fig:2}
\end{figure*}

Table \ref{tab:2} summarizes the characteristics of each subset. The train and test set are predefined with the dataset and are used unmodified following the state-of-the-art practice. The subsets are divided along two dimensions. Regarding operational conditions, all instances within $FD001$ and $FD003$ operate under a single regime at sea level, representing a setting where sensor readings reflect degradation dynamics without operational variability. In contrast, $FD002$ and $FD004$ span six distinct operating conditions. This multi-condition setting is widely regarded as more challenging, as models need to distinguish degradation-induced signal from variations caused by operational differences. Thus, emperically, the same model architectures typically show worse performance on $FD002$ and $FD004$ relative to their single-condition counterparts. Regarding fault variability, $FD001$ and $FD002$ correspond to one single fault mode, while $FD003$ and $FD004$ introduce a second fault mode corresponding to fan degradation.

\begin{table}[ht]
\footnotesize
\centering
\begin{threeparttable}
\caption{Description of the C-MAPSS subsets.}
\label{tab:2}
\begin{tabular}{lcccc}
\toprule
\textbf{Subset} & \bm{$FD001$} & \bm{$FD002$} & \bm{$FD003$} & \bm{$FD004$} \\
\midrule
Instances in training set & 100 & 260 & 100 & 249 \\
Instances in testing set & 100 & 259 & 100 & 248 \\
Fault modes & 1 & 1 & 2 & 2 \\
Operating conditions\tnote{a} & 1\tnote{b} & 6\tnote{c} & 1\tnote{b} & 6\tnote{c} \\
Max/min cycle for train & 362/128 & 378/128 & 525/145 & 543/128 \\
Max/min cycle for test & 303/31 & 367/21 & 475/38 & 486/19 \\
\bottomrule
\end{tabular}
\begin{tablenotes}
\footnotesize
\item[a] Conditions are expressed as (Altitude [ft], Mach number [--], Throttle resolver angle [$^\circ$])
\item[b] Single condition: (0, 0, 100)
\item[c] Six conditions: (0, 0, 100), (10,000, 0.25, 100), (20,000, 0.70, 100), (25,000, 0.62, 60), (35,000, 0.84, 100), (42,000, 0.84, 100)
\end{tablenotes}
\end{threeparttable}
\end{table}

The turbofan engine modeled in C-MAPSS consists of several interconnected components, as illustrated in Figure~\ref{fig:2}. Air enters through the fan and is subsequently compressed through the low-pressure compressor (LPC) and HPC. The compressed air then enters the combustion chamber (CC), where fuel is injected and ignited. The resulting high-energy exhaust drives the high-pressure turbine (HPT) and low-pressure turbine (LPT), which in turn power the compressors via concentric shafts ($N1$ and $N2$). The exhaust exits through the nozzle, generating thrust. This sequential flow path establishes thermodynamic and mechanical interdependencies among components: degradation originating in one component propagates through the system, manifesting as correlated deviations across multiple sensors.

The dataset provides 21 sensor channels per engine, measuring physical quantities including temperatures, pressures, rotational speeds, and flow rates at various locations throughout the engine. Table \ref{tab:3} details each sensor, including its symbol, physical description, units, and the component to which it is physically linked. Following established practice on C-MAPSS dataset in the literature, seven sensors, marked as grey in Table \ref{tab:3}, are excluded for RUL prediction, as these measurement are stable. The remaining 14 sensors, distributed across the Fan, LPC, HPC, CC, HPT, and LPT, constitute the input feature set.

\begin{table}[htbp]
    \footnotesize
    \centering
    \begin{threeparttable}
    \setlength{\tabcolsep}{3pt} 
    \caption{Description of the C-MAPSS sensors and component mapping.}
    \label{tab:3}
    \begin{tabular}{c l l l}
        \toprule
        \textbf{Index} & \textbf{Symbol} & \textbf{Description [Unit]} & \textbf{Component} \\
        \midrule
        \color{gray}1 & \color{gray}T2 & \color{gray}Fan Inlet Total Temp. [°R] & \color{gray}/ \\
        2 & T24 & LPC Outlet Total Temp. [°R] & LPC \\
        3 & T30 & HPC Outlet Total Temp. [°R] & HPC \\
        4 & T50 & LPT Outlet Total Temp. [°R] & LPT \\
        \color{gray}5 & \color{gray}P2 & \color{gray}Fan Inlet Pressure [psia] & \color{gray}/ \\
        \color{gray}6 & \color{gray}P15 & \color{gray}Bypass Duct Total Press. [psia] & \color{gray}/ \\
        7 & P30 & HPC Outlet Total Press. [psia] & HPC \\
        8 & Nf & Physical Fan Speed [rpm] & Fan \\
        9 & Nc & Physical Core Speed [rpm] & Core \\
        \color{gray}10 & \color{gray}epr & \color{gray}Engine Press. Ratio (P50/P2) [--] & \color{gray}/ \\
        11 & Ps30 & HPC Outlet Static Press. [psia] & HPC \\
        12 & phi & Fuel Flow / Ps30 Ratio [pps/psi] & Comb. \\
        13 & NRf & Corrected Fan Speed [rpm] & Fan \\
        14 & NRc & Corrected Core Speed [rpm] & Core \\
        15 & BPR & Bypass Ratio [--] & Fan \\
        \color{gray}16 & \color{gray}farB & \color{gray}Burner Fuel-Air Ratio [--] & \color{gray}/ \\
        17 & htBleed & Bleed Enthalpy [--] & LPC \\
        \color{gray}18 & \color{gray}Nf\_dmd & \color{gray}Demanded Fan Speed [rpm] & \color{gray}/ \\
        \color{gray}19 & \color{gray}PCNfR\_dmd & \color{gray}Demanded Corr. Fan Speed [rpm] & \color{gray}/ \\
        20 & W31 & HPT Coolant Bleed [lbm/s] & HPT \\
        21 & W32 & LPT Coolant Bleed [lbm/s] & LPT \\
        \bottomrule
    \end{tabular}%

    \vspace{0.2cm}

    \small \textit{Notes}: Rows in \color{gray}gray \color{black} indicate sensors commonly excluded. Units are reported as in the original C-MAPSS specification~\cite{frederick2007user}.

    \end{threeparttable}

\end{table}

The prediction target is the RUL, defined as the number of operational cycles until engine failure. Following existing works, a piecewise linear degradation model is adopted: RUL decreases linearly from the onset of detectable degradation but is clipped at a maximum value of 125 cycles, as early-stage operation corresponds to a healthy asset where degradation signatures are negligible while predictive accuracy in the low-RUL case carries higher importance for maintenance decision-making \cite{asif2022deep}. The input to each model is a multivariate time-series window of shape $w \times 14$, where $w$ denotes the window size in cycles, set to 30 for single-condition subsets ($FD001$ and $FD003$) and 15 for multi-condition subsets ($FD002$ and $FD004$) following optimal sizes reported in existing literature \cite{gao2024interpretable,sateesh2016deep}. Sensor readings are processed using z-score normalization with statistics computed per subset from the training set and applied to the test set. The output is a scalar RUL value, normalized to the range $[0, 1]$, where a value of 1 corresponds to the clipped maximum of 125 cycles. For training, all available windows are extracted from each trajectory using a step size of 1 to maximize data volume. For testing, consistent with the evaluation protocol in existing C-MAPSS studies, only the final window of each engine trajectory is used to predict its RUL.

\section{Case Study I: Empirical Benchmarking of Off-the-Shelf Time-Series Foundation Models} \label{sec:4}

\begin{figure*}[h]
    \includegraphics[width=1\linewidth]{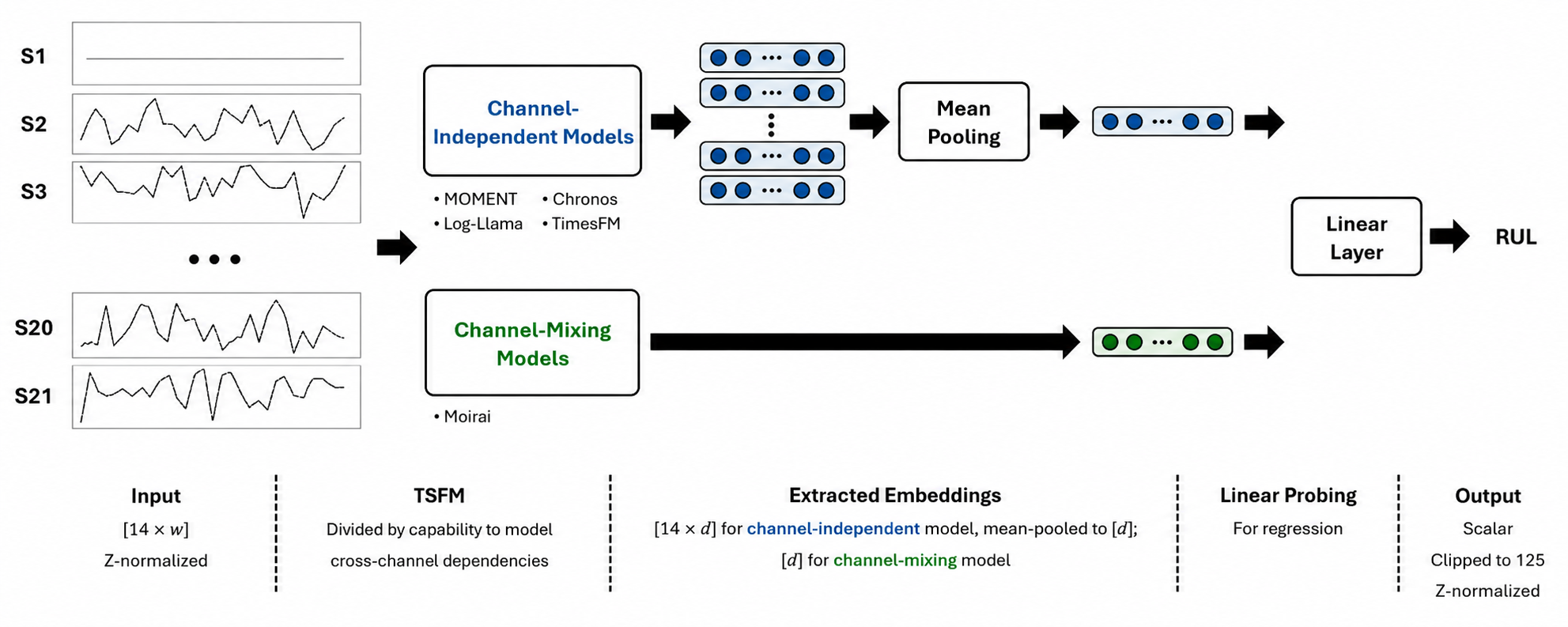}
    \centering
    \caption{Overview of the unified evaluation pipeline for benchmarking TSFMs.}
    \label{fig:3}
\end{figure*}

\subsection{Benchmark Setup} \label{sec:4.1}
All experiments employ a unified training pipeline on Google Colab with NVIDIA A100 GPU. Training is conducted using the Adam optimizer with a learning rate of $5 \times 10^{-4}$ and a batch size of 32. The loss function is mean squared error (MSE) computed on the normalized RUL targets. Performance is measured using root mean squared error (RMSE) and the NASA scoring function, as shown in Equation~\ref{eq:nasa_score}, the latter of which penalizes late predictions to reflect the higher risk associated with overestimating asset life in safety-critical systems. A ReduceLROnPlateau scheduler monitors validation RMSE, reducing the learning rate by a factor of 0.5 after 5 epochs without improvement. Model selection is carried out inside the training set by ten-fold cross-validation, with the test set of each subset held aside for reporting metrics. The folds are formed over engine trajectories rather than over windows: the units of a subset are partitioned into ten folds, and the window extraction described in Section~\ref{sec:3} is applied. Each fold trains for 25 epochs on the remaining nine folds, validates on the held-out one, and carries forward the epoch at which validation RMSE is lowest. Each of the ten resulting models is then evaluated once on the test set, and the values reported for the models proposed in this work are the mean and standard deviation over the ten folds.
\begin{equation}
    S = \sum_{i=1}^{N} s_i, \quad \text{where } s_i = 
    \begin{cases} 
    e^{-\frac{d_i}{13}} - 1 & \text{for } d_i < 0 \\
    e^{\frac{d_i}{10}} - 1 & \text{for } d_i \ge 0 
    \end{cases}
    \label{eq:nasa_score}
\end{equation}
where $d_i = \text{Predicted RUL}_i - \text{True RUL}_i$ is the error for the $i$-th sample. 

A unified linear probing strategy is employed once embeddings are extracted from each TSFM, as illustrated in Figure~\ref{fig:3}. All five models produce embeddings of shape $(batch, 14, d)$, where $d$ denotes the embedding dimension. The key distinction is in what happens during encoding. For models that cannot model cross-channel dependencies, each of the 14 channels is processed in isolation through the backbone. For the model that learned cross-channel dependencies during pretraining, all 14 channels are processed jointly through attention mechanism during encoding, allowing cross-channel information to be modeled. These embeddings then enter mean pooling and linear probing layer to generate the scalar RUL prediction. This unified path ensures that performance differences across models can be attributed to the embeddings and model characteristics themselves. Crucially, linear probing is not intended as the optimal inference strategy for deployment; rather, it serves as an efficient evaluation tool to quantify how much meaningful information the frozen embeddings encode from a performance standpoint.

The model-specific configurations for each TSFM could be found in Table~\ref{tab:4}. The following paragraphs detail these configurations, including pretrained checkpoint, input preparation, and process for embedding extraction.

\textbf{MOMENT.} The \texttt{MOMENT-1-small} checkpoint is employed as the frozen backbone. Input sequences are right-padded to the model's fixed context length of 512, with a binary mask indicating valid positions. Each sensor channel is processed independently through the encoder, and the model's embedding is retrieved as a fixed-dimensional representation of size 512. 


\textbf{Moirai.} The \texttt{moirai-1.1-R-small} checkpoint is employed the frozen backbone. Unlike other models in this study, Moirai employs an architecture via Any-variate attention, which processes all sensor channels jointly rather than independently through pretrained weights. Input sequences are left-padded to the nearest multiple of the patch size, which is set to 8, with a context length of 64, both user-defined values selected to accommodate the short window sizes characteristic of this case study. To extract embeddings, encoder's hidden states is unflattened and mean-pooled across the patch dimension. 


\textbf{Lag-Llama.} The \texttt{lag-llama.ckpt} checkpoint is employed as the frozen backbone. As a decoder-only architecture designed for causal forecasting, Lag-Llama is left-padded to a context length of 64, with a mask distinguishing valid positions from padding. Each sensor channel is processed independently with the hidden state at the final sequence position extracted as the channel embedding. The resulting embedding dimension is 144, determined by the number of attention heads. Notably, the original model's lag sequence, which encodes periodicity inductive biases, is filtered to retain only lags smaller than the context length.

\textbf{Chronos.} The \texttt{chronos-t5-small}\footnote{A subsequent version, Chronos 2, has since been released \cite{ansari2025chronos2}; however, as it became available after the experimental phase of this work was completed, the original Chronos checkpoint is used throughout.} checkpoint is employed as the frozen backbone. Each sensor channel is processed independently with hidden states of encoder extracted with a dimension of 512. To obtain a fixed-length embedding from the variable-length encoder output, mean pooling is applied over all valid token positions, weighted by the attention mask. Unlike other models, Chronos requires no explicit padding, the tokenizer naturally accommodates variable-length inputs.

\textbf{TimesFM.} The \texttt{timesfm-2.5-200m} checkpoint is employed as the frozen backbone. TimesFM is a decoder-only model that processes inputs in fixed-size patches of 32; accordingly, input sequences are left-padded to a context length of 64, with a mask indicating valid versus padded positions. Each sensor channel is processed independently with output embedding corresponding to the final patch extracted. The resulting embedding dimension is 1280, the largest among all evaluated models. 

\begin{table}[t!]
\centering
\caption{Structural and parameter comparison of TSFMs for the benchmark.}
\label{tab:4}
\vspace{2mm}
\resizebox{\textwidth}{!}{%
\begin{tabular}{@{}llllll@{}}
\toprule
\textbf{Parameter} & \textbf{MOMENT} & \textbf{Moirai} & \textbf{Lag-Llama} & \textbf{Chronos} & \textbf{TimesFM} \\
\midrule
Pretrained Checkpoint & \texttt{MOMENT-1-small} & \texttt{moirai-1.1-R-small} & \texttt{lag-llama.ckpt} & \texttt{chronos-t5-small} & \texttt{timesfm-2.5-200m} \\
Embedding Dimension   & 512                    & 384                         & 144                    & 512                       & 1,280                     \\
Context Length        & 512 (model-defined)    & 64 (user-defined)           & 64 (user-defined)      & Variable (user-defined)   & 64 (user-defined)         \\
Patch Size            & 8 (model-defined)      & 8 (user-defined)            & N/A (point-wise)       & N/A (tokenized)           & 32 (model-defined)        \\
Padding Strategy      & Right-pad              & Left-pad                    & Left-pad               & None                      & Left-pad                  \\
Frozen Parameters     & 35.3M                  & 13.8M                       & 2.4M                   & 46.2M                     & 231.3M                    \\
Multivariate     & No                 & Yes                       & No                 & No                    & No                   \\
\bottomrule
\end{tabular}%
}
\end{table}

\subsection{Benchmark Results and Discussion} \label{sec:4.2}
Table~\ref{tab:5} presents the results across all five TSFMs on the four C-MAPSS subsets. For reference, the table also includes results from state-of-the-art methods specifically designed and trained for this task.

\begin{sidewaystable}[htbp]
    \footnotesize
    \centering
    \setlength{\tabcolsep}{3pt}
    \caption{Benchmark of TSFMs on C-MAPSS dataset across different window sizes and subsets. For the five TSFMs, values are the mean $\pm$ std. over the ten folds mentioned in Section~\ref{sec:4.1}; state-of-the-art values are cited from existing works.}
    \label{tab:5}    
    \begin{tabular}{@{}ll c c c c c c c c@{}}
        \toprule
        \multirow{2}{*}{\textbf{Model}} & \multirow{2}{*}{\bm{$w$}} & \multicolumn{2}{c}{\bm{$FD001$}} & \multicolumn{2}{c}{\bm{$FD002$}} & \multicolumn{2}{c}{\bm{$FD003$}} & \multicolumn{2}{c}{\bm{$FD004$}} \\ 
        \cmidrule(lr){3-4} \cmidrule(lr){5-6} \cmidrule(lr){7-8} \cmidrule(l){9-10}
        & & RMSE & Score & RMSE & Score & RMSE & Score & RMSE & Score \\ 
        \midrule
        \multirow{2}{*}{\shortstack[l]{MOMENT \\ \footnotesize{\texttt{(MOMENT-1-small)}}}}
          & 15 & --- & --- & 49.05$\pm$0.63 & 81264$\pm$19081 & --- & --- & 49.38$\pm$0.63 & 81631$\pm$15824 \\
          & 30 & 19.13$\pm$0.15 & 1038$\pm$118 & --- & --- & 20.98$\pm$0.36 & 1116$\pm$168 & --- & --- \\
        \midrule
        \multirow{2}{*}{\shortstack[l]{Moirai \\ \footnotesize{\texttt{(moirai-1.1-R-small)}}}} 
          & 15 & --- & --- & 37.38$\pm$0.29 & 55334$\pm$8040 & --- & --- & 38.15$\pm$0.44 & 60837$\pm$9780 \\
          & 30 & 16.40$\pm$0.17 & 628$\pm$60 & --- & --- & 17.83$\pm$0.24 & 637$\pm$52 & --- & --- \\
        \midrule
        \multirow{2}{*}{\shortstack[l]{Lag-llama \\ \footnotesize{\texttt{(lag-llama.ckpt)}}}} 
          & 15 & --- & --- & 50.99$\pm$0.78 & 109430$\pm$24253 & --- & --- & 51.70$\pm$0.86 & 114530$\pm$21379 \\
          & 30 & 20.01$\pm$0.23 & 1104$\pm$77 & --- & --- & 19.69$\pm$0.31 & 794$\pm$65 & --- & --- \\
        \midrule
        \multirow{2}{*}{\shortstack[l]{Chronos \\ \footnotesize{\texttt{(chronos-t5-small)}}}}
          & 15 & --- & --- & 45.33$\pm$0.64 & 91847$\pm$20045 & --- & --- & 46.47$\pm$0.70 & 99475$\pm$23710 \\
          & 30 & 18.58$\pm$0.21 & 724$\pm$71 & --- & --- & 20.04$\pm$0.28 & 997$\pm$101 & --- & --- \\
        \midrule
        \multirow{2}{*}{\shortstack[l]{TimesFM \\ \footnotesize{\texttt{(timesfm-2.5-200m)}}}} 
          & 15 & --- & --- & 52.03$\pm$0.79 & 136853$\pm$25133 & --- & --- & 48.59$\pm$0.76 & 128796$\pm$22861 \\
          & 30 & 17.91$\pm$0.24 & 693$\pm$84 & --- & --- & 19.94$\pm$0.39 & 825$\pm$114 & --- & --- \\
        \midrule\midrule
        \textit{Comparisons (state-of-the-art)} & & & & & & & & & \\
        \addlinespace
        SVR \cite{sateesh2016deep} & 15 & 20.96 & 1382 & 42.00 & 589900 & 21.05 & 1598 & 45.35 & 37114 \\
        CNN \cite{sateesh2016deep} & 15 & 18.45 & 1287 & 30.29 & 13570 & 19.82 & 1596 & 29.16 & 7886 \\
        LSTM \cite{zheng2017long} & --- & 16.14 & 338 & 24.49 & 4450 & 16.18 & 852 & 28.17 & 5550 \\
        AGCNN \cite{liu2021life} & $18 \sim 30$ & 12.42 & 226 & 19.43 & 1492 & 13.39 & 227 & 21.50 & 3392 \\
        MCLSTM \cite{xiang2021deep} & 30 & 13.71 & 315 & --- & --- & --- & --- & 23.81 & 4826 \\
        BiGRU-TSAM\textsuperscript{a} \cite{zhang2022prediction} & $10 \sim 30$ & 12.56 & 213 & 18.94 & 2264 & 12.45 & 233 & 20.47 & 3610 \\
        DRLRULe\textsuperscript{b} \cite{hu2023remaining} & $18 \sim 30$ & 12.17 & 208 & 16.28 & 1437 & 13.09 & 226 & 18.87 & 1726 \\
        \bottomrule
        \multicolumn{10}{l}{\footnotesize \textsuperscript{a} Apart from the selected 14 sensors, for complex subsets ($FD002$, $FD004$), the model explicitly includes 3 operating variables and} \\
        \multicolumn{10}{l}{\footnotesize \quad\ 7 sensors typically excluded by many existing works, totaling 24 input variables.} \\
        \multicolumn{10}{l}{\footnotesize \textsuperscript{b} Apart from the 14 sensors, for complex subsets ($FD002$, $FD004$), the model implicitly uses 3 operating variables, corresponding to} \\
        \multicolumn{10}{l}{\footnotesize \quad\ 6 conditions, for specific normalization, while restricting the network input to the 14 selected sensor variables.} \\
    \end{tabular}
\end{sidewaystable}
\textbf{Subset-dependent performance differences and computational efficiency.} The performance differences between off-the-shelf TSFMs and state-of-the-art methods is not uniform; rather, it is strongly conditioned on operational complexity. On single-condition subsets, TSFMs demonstrate similar performance relative to some state-of-the-art models, with the strongest TSFM results approaching or matching state-of-the-art baselines such as LSTM. On complex subsets spanning six operating conditions, however, the difference increases substantially across all TSFMs. This subset-dependent pattern suggests that current TSFMs struggle to distinguish degradation-related signal from variations attribute to operational changes---a capability that state-of-the-art architectures acquire through explicit design choices such as condition-specific normalization or context-aware inputs. Nevertheless, this performance difference does not make TSFMs irrelevant for industrial applications. Beyond the generalization offered by broad pre-training, the linear probing approach employed here requires only a small number of trainable parameters while keeping the backbone entirely frozen, in contrast to state-of-the-art models that are trained end-to-end on task-specific data. This efficiency becomes particularly valuable in digital twin environments where multiple downstream tasks must be supported simultaneously without maintaining a dedicated model for each.

\textbf{Predictive advantage of multivariate architectures.} Among all evaluated TSFMs, Moirai exhibits consistently better performance across all four subsets, and the marginal improvements over models without cross-channel dependecies learned is especially large on the complex multi-condition subsets. This comes directly from its Any-variate Attention mechanism, which processes all 14 sensor channels jointly during encoding rather than independently. By allowing cross-channel information to flow during the forward pass, Moirai can capture the correlated sensor deviations that propagate across engine components when degradation originates in one component. Notably, this performance is achieved with fewer than 400 trainable parameters, the smallest count among all evaluated models, reflecting the minimal adaptation required when the backbone inherently aligns with the multivariate nature of the downstream task.

\textbf{Performance divergence across univariate models.} Within models without cross-channel dependecies learned, the simple and complex subsets show highly different patterns. On simple subsets, most univariate models cluster within a narrow performance band, suggesting that single-condition scenarios do not strongly differentiate architectural characteristics. On complex subsets, the performances across models are no longer close. Chronos separates from the group as the strongest model. Chronos' relative robustness on complex subsets can be attributed to its quantization mechanism: by mapping continuous values into a fixed discrete vocabulary, the representation becomes partially insensitive to the magnitude shifts induced by different operating conditions, providing an implicit form of scale invariance that patch-based models lack \cite{masserano2024enhancing}. Despite these within-group differences, all four models largely fell short of state-of-the-art performance on complex subsets.

\textbf{Cross-channel dependecies as the discriminating factor.} The results show a pattern: the modeling of cross-channel dependecies during encoding is a prominent driver of performance on this task, particularly under complex operational conditions. Yet this raises a deeper question than simply which backbone to select. The cross-channel dependecies encoded in Moirai's pretrained weights originates from massive, heterogeneous pre-training corpora spanning domains that are unrelated to turbofan degradation; whether these patterns are genuinely informative for this task, or whether cross-channel dependencies, learned through task-specific adaptation or derived from digital twin, contribute information that pretrained weights cannot provide, remains entirely open. The same question applies to those TSFMs with pretrained cross-channel dependecies missing. Understanding the individual contribution of each source, through pretrained weights, through target-task adaptation, and from digital twin, and whether they are complementary or redundant is what motivates the fusion of TSFM and digital twin representation proposed in Section~\ref{sec:5}, and the continuation of this case study as an ablation study shown in Section~\ref{sec:6}. 

\section{Using Asset Topology from the Digital Twin to Inform Time-Series Foundation Models}\label{sec:5}

\subsection{Encoding Asset Topology via Hierarchical Message Passing} \label{sec:5.1}

Section~\ref{sec:2} established that graph structures serve as a common form of digital twin representation across industrial domains, encoding topological relationships that inform downstream reasoning. Here we formalize one such representation, a hierarchical topological graph capturing the sensor-component-asset structure that a digital twin commonly stores among its information models, and describe how cross-channel relationships are encoded from this structure via message passing.

We define a topological graph $\mathcal{G} = (\mathcal{V}, \mathcal{E})$ where the node set $\mathcal{V} = \mathcal{V}_s \cup \mathcal{V}_c \cup \mathcal{V}_a$ is partitioned into three disjoint subsets. $\mathcal{V}_s = \{s_1, \ldots, s_n\}$ denotes $n$ sensor nodes at the instrumentation level, $\mathcal{V}_c = \{c_1, \ldots, c_m\}$ denotes $m$ component nodes, and $\mathcal{V}_a = \{a\}$ is the single asset node. The edge set $\mathcal{E}$ is similarly partitioned: inter-level edge sets $\mathcal{E}_{s \to c} \subseteq \mathcal{V}_s \times \mathcal{V}_c$ and $\mathcal{E}_{c \to a} \subseteq \mathcal{V}_c \times \mathcal{V}_a$ are directed, capturing hierarchical aggregation from sensors to components and from components to the asset, respectively; the intra-level edge set $\mathcal{E}_{c \leftrightarrow c} \subseteq \mathcal{V}_c \times \mathcal{V}_c$ is undirected, reflecting the symmetric nature of physical interactions between subsystems. 

\begin{figure*}[b!]
    \includegraphics[width=1\linewidth]{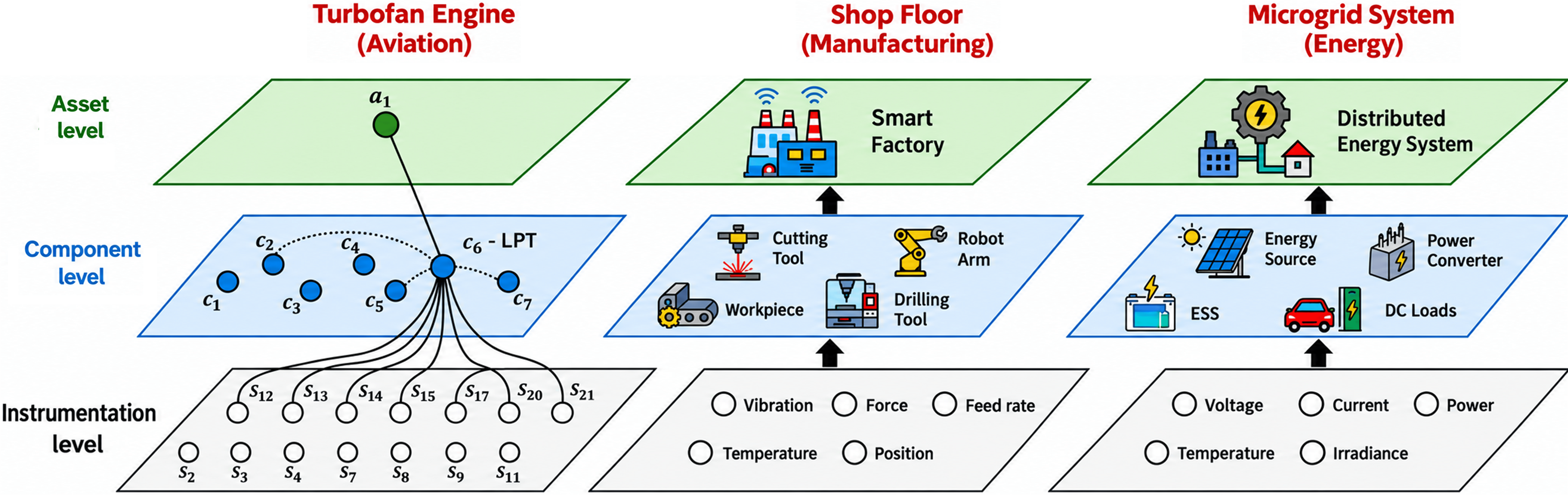}
    \centering
    \caption{Hierarchical topological graph illustrated for aviation (left), manufacturing (center), and energy (right) domains. The aviation instantiation corresponds to the C-MAPSS turbofan with LPT-associated ($c_6$) edges highlighted; manufacturing and energy columns depict schematic representations of typical component and sensor types from respective fields \cite{jia2023twin, namita2023twin}.}
    \label{fig:4}
\end{figure*}

Connectivity is encoded through adjacency matrices: $\mathbf{A}_{s \to c} \in \{0,1\}^{n \times m}$ for sensor-to-component relationships, $\mathbf{A}_{c \leftrightarrow c} \in \{0,1\}^{m \times m}$ for component-level relationships, and $\mathbf{A}_{c \to a} \in \{0,1\}^{m \times 1}$ for component-to-asset relationships. The rules governing edge construction are inherently domain-dependent, reflecting material flow in manufacturing, electrical connectivity in energy systems, or thermodynamic flow paths in aerospace, but in all cases, the topology is derived from the asset structure the digital twin already records among its information models. Figure~\ref{fig:4} illustrates this hierarchical structure across three domains, with the aviation column corresponding to the C-MAPSS turbofan instantiation used throughout this study.
 
Given this graph structure, cross-channel relationships are encoded in hierarchical message passing over $\mathcal{G}$. Algorithm~\ref{alg:1} details this process. In the algorithm, $x_i \in \mathbb{R}^w$ denotes the $i$-th row of $\mathbf{X}$; $\sigma$ is a nonlinear activation function; and $\mathbf{W}^{(s)}$, $\mathbf{b}^{(s)}$, $\mathbf{W}^{(\text{agg})}$, $\mathbf{W}^{(\text{lat})}$, $\mathbf{W}^{(\text{asset})}$ are learnable parameter matrices and bias vectors. The resulting component-level embeddings $\mathbf{H}_c \in \mathbb{R}^{m \times d}$ (where $d$ is the shared embedding dimension used throughout) constitute the digital twin-derived cross-channel dependecies used in this work: each embedding encapsulates the component's learned embedding, the aggregated state of its associated sensors, and contextual information from adjacent components.
 
\begin{algorithm}[h]
\caption{Hierarchical message passing over $\mathcal{G}$}
\label{alg:1}
\begin{algorithmic}[1]
\REQUIRE Sensor observations $\mathbf{X} \in \mathbb{R}^{n \times w}$, adjacency matrices $\mathbf{A}_{s \rightarrow c}$, $\mathbf{A}_{c \leftrightarrow c}$, $\mathbf{A}_{c \rightarrow a}$, learnable component embeddings $\{\mathbf{e}_j^{(c)}\}_{j=1}^{m}$
\ENSURE Component embeddings $\mathbf{H}_c \in \mathbb{R}^{m \times d}$, asset embedding $\mathbf{h}^{(a)} \in \mathbb{R}^d$
 
\STATE \textit{Sensor-level projection:}
\FOR{each sensor $s_i \in \mathcal{V}_s$}
    \STATE $\mathbf{h}_i^{(s)} \leftarrow \sigma\!\left(\mathbf{W}^{(s)} \mathbf{x}_i + \mathbf{b}^{(s)}\right)$
\ENDFOR
 
\STATE \textit{Sensor $\rightarrow$ Component aggregation:}
\FOR{each component $c_j \in \mathcal{V}_c$}
    \STATE $\mathcal{N}_j^{(s)} \leftarrow \{i : [\mathbf{A}_{s \rightarrow c}]_{ij} = 1\}$
    \STATE $\mathbf{m}_j^{(s)} \leftarrow \frac{1}{|\mathcal{N}_j^{(s)}|} \sum_{i \in \mathcal{N}_j^{(s)}} \mathbf{h}_i^{(s)}$
    \STATE $\mathbf{h}_j^{(c)} \leftarrow \sigma\!\left(\mathbf{W}^{(\mathrm{agg})} \cdot \mathrm{CONCAT}\!\left[\mathbf{e}_j^{(c)},\, \mathbf{m}_j^{(s)}\right]\right)$
\ENDFOR
 
\STATE \textit{Component $\leftrightarrow$ Component lateral exchange:}
\FOR{each component $c_j \in \mathcal{V}_c$}
    \STATE $\mathcal{N}_j^{(c)} \leftarrow \{k : [\mathbf{A}_{c \leftrightarrow c}]_{jk} = 1\}$
    \IF{$|\mathcal{N}_j^{(c)}| > 0$}
        \STATE $\mathbf{m}_j^{(c)} \leftarrow \frac{1}{|\mathcal{N}_j^{(c)}|} \sum_{k \in \mathcal{N}_j^{(c)}} \mathbf{h}_k^{(c)}$
        \STATE $\tilde{\mathbf{h}}_j^{(c)} \leftarrow \sigma\!\left(\mathbf{W}^{(\mathrm{lat})} \cdot \mathrm{CONCAT}\!\left[\mathbf{h}_j^{(c)},\, \mathbf{m}_j^{(c)}\right]\right)$
    \ELSE
        \STATE $\tilde{\mathbf{h}}_j^{(c)} \leftarrow \mathbf{h}_j^{(c)}$
    \ENDIF
\ENDFOR
 
\STATE \textit{Component $\rightarrow$ Asset aggregation:}
\STATE $\mathbf{h}^{(a)} \leftarrow \sigma\!\left(\mathbf{W}^{(\mathrm{asset})} \cdot \frac{1}{m} \sum_{j=1}^{m} \tilde{\mathbf{h}}_j^{(c)}\right)$
\RETURN $\mathbf{H}_c = [\tilde{\mathbf{h}}_1^{(c)};\, \ldots;\, \tilde{\mathbf{h}}_m^{(c)}]$, $\mathbf{h}^{(a)}$
\end{algorithmic}
\end{algorithm}

\subsection{Informing TSFM Embeddings via Topology-Informed Attention} \label{sec:5.2}
With component-level embeddings extracted via hierarchical message passing, the next step is to integrate this digital twin-derived cross-channel dependecies with the sensor-level embeddings produced by the TSFM backbone. Cross-attention allows sensor-level TSFM embeddings serve as queries that attend to component-level embeddings, allowing each sensor's temporal representation to be informed by the structural context of the subsystem it monitors and those to which it is physically connected.

However, standard cross-attention allows unconstrained all-to-all interactions, which contradicts the sparse, local nature of physical systems where components primarily influence their immediate neighbors. To enforce locality, a binary mask $\mathbf{M} \in \mathbb{R}^{n \times m}$ is derived from the adjacency matrices defined in Section~\ref{sec:5.1}. For each sensor $s_i$ associated with component $c_j$ via $\mathbf{A}_{s \rightarrow c}$, the mask allows attention not only to $c_j$ but also to all components $c_k$ that are one-hop neighbors of $c_j$ according to $\mathbf{A}_{c \leftrightarrow c}$. All other positions receive $-\infty$, yielding effectively zero attention weight after softmax. The resulting informed embeddings $\mathbf{H}^{(\mathrm{informed})} \in \mathbb{R}^{n \times d}$ retain the temporal information captured by the frozen TSFM while incorporating digital twin-derived cross-channel dependencies constrained by the physical topology. 

Importantly, this fusion mechanism is agnostic to the TSFM backbone capability to model cross-channel dependencies or not. For a TSFM without cross-channel dependencies learned through pretraining like MOMENT, each sensor embedding $\mathbf{h}_i^{(\mathrm{TSFM})}$ is generated independently with no cross-channel information encoded during the forward pass; the masked cross-attention then serves as the mechanism through which sensor-level representations gain awareness of the physical system's relational structure. On the other hand, for a TSFM with cross-channel dependencies learned through pretraining like Moirai, sensor embeddings already carry pretrained cross-channel dependencies learned from large-scale corpora via attention mechanism; the masked cross-attention supplements this with an asset-specific source of cross-channel dependencies derived as digital twin representation. 

\section{Case Study II: Ablation Study on Cross-Channel Dependency Sources} \label{sec:6}
\subsection{Ablation Study Setup} \label{sec:6.1}
The fusion approach proposed in Section~\ref{sec:5} allows the integration of multiple sources of information about cross-channel dependencies. However, the individual contribution of each source to downstream task performance remains unclear. This directly relates to the central question motivating this work: how much cross-channel information does each source---through pretrained weights, through target-task adaptation, and from digital twin---provide to general-purpose TSFMs, and to what extent are their contributions complementary? The third of these sources, however, has been argued at the theoretical level as a position: existing work points out that digital twins offer physically grounded representations that explicitly encode domain knowledge and preserve the continuous nature of real-world processes that TSFMs often do not hold \cite{shen2025position}. That position applies to digital twin representations in general; it has not been empirically shown for a specific one, such as representations derived from connectivity among components and sensors described in Section~\ref{sec:5}. To address these questions empirically, this section presents an ablation study that first establishes what the connectivity contributes, and then systematically isolates three sources, revealing both their individual effects

The representation formalized in Section~\ref{sec:5} is derived from the sensor-component structure of Table~\ref{tab:3} and Figure~\ref{fig:4}, referred to below as the ground truth structure. To establish that digital twin representations derived from connectivity among components and sensors contain meaningful cross-channel dependencies \cite{chan2026structural}, they are compared against two sets of graphs of the same size. Under incorrect component coupling, the connections among components are redistributed, so each component is coupled to the same number of others but not to the right ones. Under incorrect sensor-component coupling, the component structure is left as it is and the sensor assignments are shuffled, so each sensor is grounded to the wrong component. Both keep the number of connections and the number of neighbors at every node. All three are run in the GNN-alone configuration, which bypasses the TSFM backbone, so the effect of the connections is measured on its own. The structure drawn for each fold is listed in \ref{app:2}.

The three ablation dimensions correspond to distinct stages in the processing pipeline where cross-channel dependecies can be modeled. Cross-channel dependencies learned through pretraining (\textbf{Pretrain} in Table~\ref{tab:7}) captures whether the TSFM backbone leverages cross-channel attention learned during pre-training on diverse time-series corpora. This dimension controls whether all 14 sensor channels are processed jointly through the encoder's attention mechanism, allowing the model to apply its pretrained understanding of inter-channel dependencies. Cross-channel dependencies learned through target-task adaptation (\textbf{Target-Task} in Table~\ref{tab:7}) captures whether channel-mixing layer are learned specifically from the C-MAPSS training data during adaptation. This dimension controls whether a learnable depthwise convolution explicitly models how the 14 sensors co-vary for RUL prediction on this particular asset. Cross-channel dependencies derived from digital twin (\textbf{Digital Twin Derived} in Table~\ref{tab:7}) capture whether physics-informed representations derived from the turbofan's hierarchical topology are injected into sensor embeddings via cross-attention. This dimension has three variants: masked fusion, which constrains cross-attention to physically connected sensor-component pairs; unmasked fusion, which allows all-to-all sensor-component interactions; and no fusion, which bypasses the topological graph entirely, leaving the TSFM to predict without any digital twin input.

The training setups follow those established in Section~\ref{sec:4.1}, with identical hyperparameters, data preprocessing, and computational environment. The TSFM backbones remain frozen throughout: \texttt{MOMENT-1-small} for MOMENT experiments and \texttt{Moirai-1.1-R-small} for Moirai experiments, with embedding dimensions $d=512$ and $d=384$ respectively. Performance is evaluated using three metrics. RMSE and the NASA scoring function (Score, Equation~\ref{eq:nasa_score}) follow the same definitions used in Section~\ref{sec:4}. Normalized RMSE (NRMSE) is introduced here to enable comparison across subsets with differing RUL ranges, computed as:
\begin{equation}
\text{NRMSE} = \frac{\text{RMSE}}{\max(\text{RUL}_{\text{test}}) - \min(\text{RUL}_{\text{test}})}
\end{equation}
where the normalization range is determined from the actual RUL values observed in each subset's test set. When aggregating across subsets, a weighted average NRMSE is computed, where weights are proportional to the number of test instances in each subset to account for their differing sizes. Each configuration is executed under the ten-fold protocol defined in Section~\ref{sec:4.1}, and results are reported as the mean ± standard deviation (std.) over the ten folds.

\subsection{Ablation Results and Discussion} \label{sec:6.2}

The following analysis focuses on NRMSE as the primary metric, as it enables comparison across subsets. In the meantime, both tables below show RMSE, NRMSE and the NASA score.

Table~\ref{tab:6} presents the GNN-alone configuration under the ground truth sensor-component structure and under the two alternatives defined in Section~\ref{sec:6.1}, across the four C-MAPSS subsets.
 
\textbf{The ground truth connectivity carries health-related information, though its contribution is limited and concentrated where operating conditions vary.} Here we test whether the connectivity among components and sensors carries health-related information, and so is a source from which a digital twin representation can be derived. The three settings hold the trainable parameter count fixed at 3,649, and hold fixed how many values are averaged at each node, so any difference between them can be attributed to which elements are treated as connected. Replacing the ground truth structure degrades performance, and the size of the effect is limited: across the four subsets, the weighted average NRMSE rises from 0.1774 under the ground truth structure to 0.1810 under incorrect component coupling and 0.1895 under incorrect sensor-component coupling, increases of 2.0\% and 6.8\% respectively. Within this effect, most of the information lies in the sensor grounding: incorrect sensor-component coupling degrades performance more than incorrect component coupling on the single-condition and the multi-condition subsets alike, indicating that the assignment of a sensor to its component carries more information than the link between two components. Both incorrect structures also degrade performance less under a single operating regime than under six. Under a single operating regime, the sensors co-vary in a stable way, so a graph specifying which sensors are attached to the same component largely duplicates information already available in the readings, and an incorrect structure degrades performance only slightly. When six regimes are interleaved, the correlations among raw readings change with the operating point, and the structure provides a regime-independent specification of which sensors are attached to the same component. This is the information required to separate degradation from operational variation, which is the distinction that TSFMs were shown to struggle with in Section~\ref{sec:4.2}. The connectivity that the digital twin records between sensors and components therefore carries information about remaining useful life that can potentially inform TSFMs \cite{djeziri2018hybrid}, though it accounts for a limited share of the GNN's predictive performance, which rests primarily on the mapping learned from the sensor readings themselves.

\begin{table}[t!]
\small
\centering
\caption{Performance (mean \protect\mbox{$\pm$} std.) of the GNN-alone configuration under the ground truth sensor-component structure and under the two alternatives, defined in Section~\protect\mbox{\ref{sec:6.1}}, across C-MAPSS subsets.}
\label{tab:6}
\resizebox{\textwidth}{!}{%
\begin{tabular}{>{\raggedright\arraybackslash}p{0.24\textwidth}|cccc}
\hline
\textbf{Asset structure} & \textbf{FD001 ($w=30$)} & \textbf{FD002 ($w=15$)} & \textbf{FD003 ($w=30$)} & \textbf{FD004 ($w=15$)} \\
\hline
\multirow{3}{0.24\textwidth}{Ground truth} & RMSE: 13.12±0.49 & RMSE: 24.08±1.36 & RMSE: 13.37±0.85 & RMSE: 24.31±0.23 \\
 & NRMSE: 0.1112±0.0042 & NRMSE: 0.2024±0.0114 & NRMSE: 0.1124±0.0071 & NRMSE: 0.2043±0.0019 \\
 & Score: 269±39 & Score: 8071±2759 & Score: 349±102 & Score: 9167±2531 \\
\hline
\multirow{3}{0.24\textwidth}{Incorrect component coupling} & RMSE: 13.18±0.80 & RMSE: 24.57±1.12 & RMSE: 13.26±0.88 & RMSE: 25.04±0.34 \\
 & NRMSE: 0.1117±0.0068 & NRMSE: 0.2064±0.0094 & NRMSE: 0.1114±0.0074 & NRMSE: 0.2104±0.0029 \\
 & Score: 281±47 & Score: 8573±2102 & Score: 319±71 & Score: 9027±2925 \\
\hline
\multirow{3}{0.24\textwidth}{Incorrect sensor-component coupling} & RMSE: 13.45±0.39 & RMSE: 25.80±0.88 & RMSE: 13.79±0.55 & RMSE: 26.32±0.60 \\
 & NRMSE: 0.1140±0.0033 & NRMSE: 0.2168±0.0074 & NRMSE: 0.1159±0.0046 & NRMSE: 0.2212±0.0050 \\
 & Score: 285±31 & Score: 10183±2874 & Score: 355±34 & Score: 13783±3784 \\
\hline
\end{tabular}%
}
\vspace{0.2cm}
\end{table}

Table~\ref{tab:7} presents the complete ablation results for both MOMENT and Moirai across all configurations and C-MAPSS subsets.

\begin{table}[p]
\centering
\caption{Performance (mean $\pm$ std.) of MOMENT and Moirai configurations across C-MAPSS subsets.}
\label{tab:7}
\resizebox{\textwidth}{!}{%
\begin{tabular}{ccc|cccc}
\hline
\textbf{Pretrain} & \textbf{\begin{tabular}{@{}c@{}} Target \\ Task \end{tabular}} & \textbf{\begin{tabular}{@{}c@{}} Digital Twin \\ Derived \end{tabular}} & \textbf{FD001 ($w=30$)} & \textbf{FD002 ($w=15$)} & \textbf{FD003 ($w=30$)} & \textbf{FD004 ($w=15$)} \\
\hline
\multicolumn{7}{c}{\textit{MOMENT}} \\
\hline
 &  & & RMSE: 12.69±0.78 & RMSE: 23.17±0.96 & RMSE: 12.95±1.04 & RMSE: 24.39±0.53 \\
\xmark & \cmark & \cmark & NRMSE: 0.1075±0.0066 & NRMSE: 0.1947±0.0081 & NRMSE: 0.1088±0.0087 & NRMSE: 0.2050±0.0045 \\
 &  & & Score: 257±54 & Score: 5980±1585 & Score: 385±130 & Score: 7541±2238 \\
\rowcolor{gray!20}  &  & & RMSE: 12.94±0.52 & RMSE: 24.74±1.16 & RMSE: 13.10±0.74 & RMSE: 25.62±0.90 \\
\rowcolor{gray!20} \xmark & \cmark & \hmark & NRMSE: 0.1097±0.0044 & NRMSE: 0.2079±0.0098 & NRMSE: 0.1101±0.0062 & NRMSE: 0.2153±0.0075 \\
\rowcolor{gray!20}  &  & & Score: 269±41 & Score: 7298±1927 & Score: 359±82 & Score: 10039±3115 \\
 &  & & RMSE: 17.21±0.45 & RMSE: 47.35±0.51 & RMSE: 17.97±0.48 & RMSE: 48.16±0.55 \\
\xmark & \cmark & \xmark & NRMSE: 0.1459±0.0038 & NRMSE: 0.3979±0.0043 & NRMSE: 0.1510±0.0041 & NRMSE: 0.4047±0.0046 \\
 &  & & Score: 673±107 & Score: 83129±16091 & Score: 1229±395 & Score: 83935±13443 \\
\rowcolor{gray!20}  &  & & RMSE: 13.06±0.75 & RMSE: 23.48±0.83 & RMSE: 13.24±1.10 & RMSE: 24.46±1.15 \\
\rowcolor{gray!20} \xmark & \xmark & \cmark & NRMSE: 0.1107±0.0064 & NRMSE: 0.1973±0.0070 & NRMSE: 0.1113±0.0092 & NRMSE: 0.2055±0.0097 \\
\rowcolor{gray!20}  &  & & Score: 269±54 & Score: 5829±1343 & Score: 355±119 & Score: 8549±2426 \\
 &  & & RMSE: 13.45±0.94 & RMSE: 24.57±0.88 & RMSE: 13.12±0.66 & RMSE: 25.33±1.11 \\
\xmark & \xmark & \hmark & NRMSE: 0.1140±0.0080 & NRMSE: 0.2064±0.0074 & NRMSE: 0.1103±0.0056 & NRMSE: 0.2129±0.0093 \\
 &  & & Score: 265±58 & Score: 8345±2385 & Score: 337±82 & Score: 10983±4743 \\
\rowcolor{gray!20}  &  & & RMSE: 19.13±0.15 & RMSE: 49.05±0.63 & RMSE: 20.98±0.36 & RMSE: 49.38±0.63 \\
\rowcolor{gray!20} \xmark & \xmark & \xmark & NRMSE: 0.1621±0.0013 & NRMSE: 0.4122±0.0053 & NRMSE: 0.1763±0.0030 & NRMSE: 0.4150±0.0053 \\
\rowcolor{gray!20}  &  & & Score: 1038±118 & Score: 81264±19081 & Score: 1116±168 & Score: 81631±15824 \\
\hline
\multicolumn{7}{c}{\textit{Moirai}} \\
\hline
 &  & & RMSE: 11.58±0.62 & RMSE: 21.15±0.88 & RMSE: 11.77±0.48 & RMSE: 22.89±0.60 \\
\cmark & \cmark & \cmark & NRMSE: 0.0982±0.0052 & NRMSE: 0.1778±0.0074 & NRMSE: 0.0989±0.0040 & NRMSE: 0.1924±0.0050 \\
 &  & & Score: 215±34 & Score: 4619±1525 & Score: 272±52 & Score: 5301±2727 \\
\rowcolor{gray!20}  &  & & RMSE: 12.33±0.63 & RMSE: 21.96±0.78 & RMSE: 12.11±0.73 & RMSE: 23.56±0.94 \\
\rowcolor{gray!20} \cmark & \cmark & \hmark & NRMSE: 0.1045±0.0053 & NRMSE: 0.1845±0.0066 & NRMSE: 0.1017±0.0062 & NRMSE: 0.1980±0.0079 \\
\rowcolor{gray!20}  &  & & Score: 237±42 & Score: 5116±1873 & Score: 278±75 & Score: 9099±5545 \\
 &  & & RMSE: 14.81±0.44 & RMSE: 35.50±0.43 & RMSE: 14.03±0.35 & RMSE: 35.71±0.52 \\
\cmark & \cmark & \xmark & NRMSE: 0.1255±0.0037 & NRMSE: 0.2983±0.0036 & NRMSE: 0.1179±0.0029 & NRMSE: 0.3001±0.0044 \\
 &  & & Score: 366±34 & Score: 58067±11565 & Score: 374±49 & Score: 74936±16196 \\
\rowcolor{gray!20}  &  & & RMSE: 12.03±0.58 & RMSE: 21.25±0.87 & RMSE: 12.14±0.94 & RMSE: 23.05±1.05 \\
\rowcolor{gray!20} \cmark & \xmark & \cmark & NRMSE: 0.1020±0.0049 & NRMSE: 0.1786±0.0073 & NRMSE: 0.1020±0.0079 & NRMSE: 0.1937±0.0088 \\
\rowcolor{gray!20}  &  & & Score: 245±43 & Score: 4844±2392 & Score: 289±110 & Score: 5912±2192 \\
 &  & & RMSE: 12.44±0.65 & RMSE: 22.14±1.04 & RMSE: 12.23±1.47 & RMSE: 23.81±0.64 \\
\cmark & \xmark & \hmark & NRMSE: 0.1054±0.0055 & NRMSE: 0.1860±0.0088 & NRMSE: 0.1028±0.0123 & NRMSE: 0.2001±0.0054 \\
 &  & & Score: 254±59 & Score: 4938±895 & Score: 290±141 & Score: 7352±2755 \\
\rowcolor{gray!20}  &  & & RMSE: 16.40±0.17 & RMSE: 37.38±0.29 & RMSE: 17.83±0.24 & RMSE: 38.15±0.44 \\
\rowcolor{gray!20} \cmark & \xmark & \xmark & NRMSE: 0.1390±0.0014 & NRMSE: 0.3142±0.0025 & NRMSE: 0.1498±0.0020 & NRMSE: 0.3206±0.0037 \\
\rowcolor{gray!20}  &  & & Score: 628±60 & Score: 55334±8040 & Score: 637±52 & Score: 60837±9780 \\
\hline
 &  & & RMSE: 12.32±0.74 & RMSE: 22.36±0.61 & RMSE: 12.31±1.20 & RMSE: 24.00±0.79 \\
\xmark & \cmark & \cmark & NRMSE: 0.1044±0.0062 & NRMSE: 0.1879±0.0052 & NRMSE: 0.1034±0.0101 & NRMSE: 0.2017±0.0067 \\
 &  & & Score: 247±37 & Score: 5166±1554 & Score: 292±97 & Score: 5622±1133 \\
\rowcolor{gray!20}  &  & & RMSE: 12.83±1.02 & RMSE: 22.81±0.58 & RMSE: 12.40±0.82 & RMSE: 24.40±0.74 \\
\rowcolor{gray!20} \xmark & \cmark & \hmark & NRMSE: 0.1087±0.0087 & NRMSE: 0.1917±0.0049 & NRMSE: 0.1042±0.0069 & NRMSE: 0.2050±0.0062 \\
\rowcolor{gray!20}  &  & & Score: 271±77 & Score: 5524±1730 & Score: 313±101 & Score: 7983±2690 \\
 &  & & RMSE: 14.94±0.55 & RMSE: 39.07±0.42 & RMSE: 13.99±0.57 & RMSE: 39.93±0.65 \\
\xmark & \cmark & \xmark & NRMSE: 0.1266±0.0046 & NRMSE: 0.3284±0.0035 & NRMSE: 0.1175±0.0048 & NRMSE: 0.3355±0.0055 \\
 &  & & Score: 394±78 & Score: 83453±12469 & Score: 327±51 & Score: 93337±22347 \\
\rowcolor{gray!20}  &  & & RMSE: 12.76±0.55 & RMSE: 22.66±1.01 & RMSE: 12.78±0.74 & RMSE: 24.31±1.48 \\
\rowcolor{gray!20} \xmark & \xmark & \cmark & NRMSE: 0.1082±0.0047 & NRMSE: 0.1905±0.0084 & NRMSE: 0.1074±0.0062 & NRMSE: 0.2043±0.0124 \\
\rowcolor{gray!20}  &  & & Score: 257±48 & Score: 6139±2784 & Score: 335±69 & Score: 7600±3419 \\
 &  & & RMSE: 13.09±0.47 & RMSE: 23.07±1.24 & RMSE: 13.06±0.75 & RMSE: 25.06±0.71 \\
\xmark & \xmark & \hmark & NRMSE: 0.1109±0.0040 & NRMSE: 0.1939±0.0104 & NRMSE: 0.1097±0.0063 & NRMSE: 0.2106±0.0059 \\
 &  & & Score: 251±40 & Score: 6009±2605 & Score: 350±71 & Score: 8747±3198 \\
\rowcolor{gray!20}  &  & & RMSE: 17.22±0.10 & RMSE: 41.26±0.36 & RMSE: 18.40±0.21 & RMSE: 41.87±0.47 \\
\rowcolor{gray!20} \xmark & \xmark & \xmark & NRMSE: 0.1459±0.0009 & NRMSE: 0.3467±0.0030 & NRMSE: 0.1546±0.0017 & NRMSE: 0.3519±0.0039 \\
\rowcolor{gray!20}  &  & & Score: 708±60 & Score: 74989±12344 & Score: 739±28 & Score: 66184±10573 \\
\hline
\multicolumn{3}{c|}{\multirow{3}{*}{\begin{tabular}{@{}c@{}} GNN using digital twin- \\ derived representations \end{tabular}}} & RMSE: 13.12±0.49 & RMSE: 24.08±1.36 & RMSE: 13.37±0.85 & RMSE: 24.31±0.23 \\
\multicolumn{3}{c|}{} & NRMSE: 0.1112±0.0042 & NRMSE: 0.2024±0.0114 & NRMSE: 0.1124±0.0071 & NRMSE: 0.2043±0.0019 \\
\multicolumn{3}{c|}{} & Score: 269±39 & Score: 8071±2759 & Score: 349±102 & Score: 9167±2531 \\
\hline
\end{tabular}%
}

\vspace{0.2cm}
\small \textit{Notes}: Enabled (\cmark), Disabled (\xmark), Unmasked (\hmark) (digital twin fusion only). 

\end{table}

\begin{figure}[t!]
    \includegraphics[width=.6\linewidth]{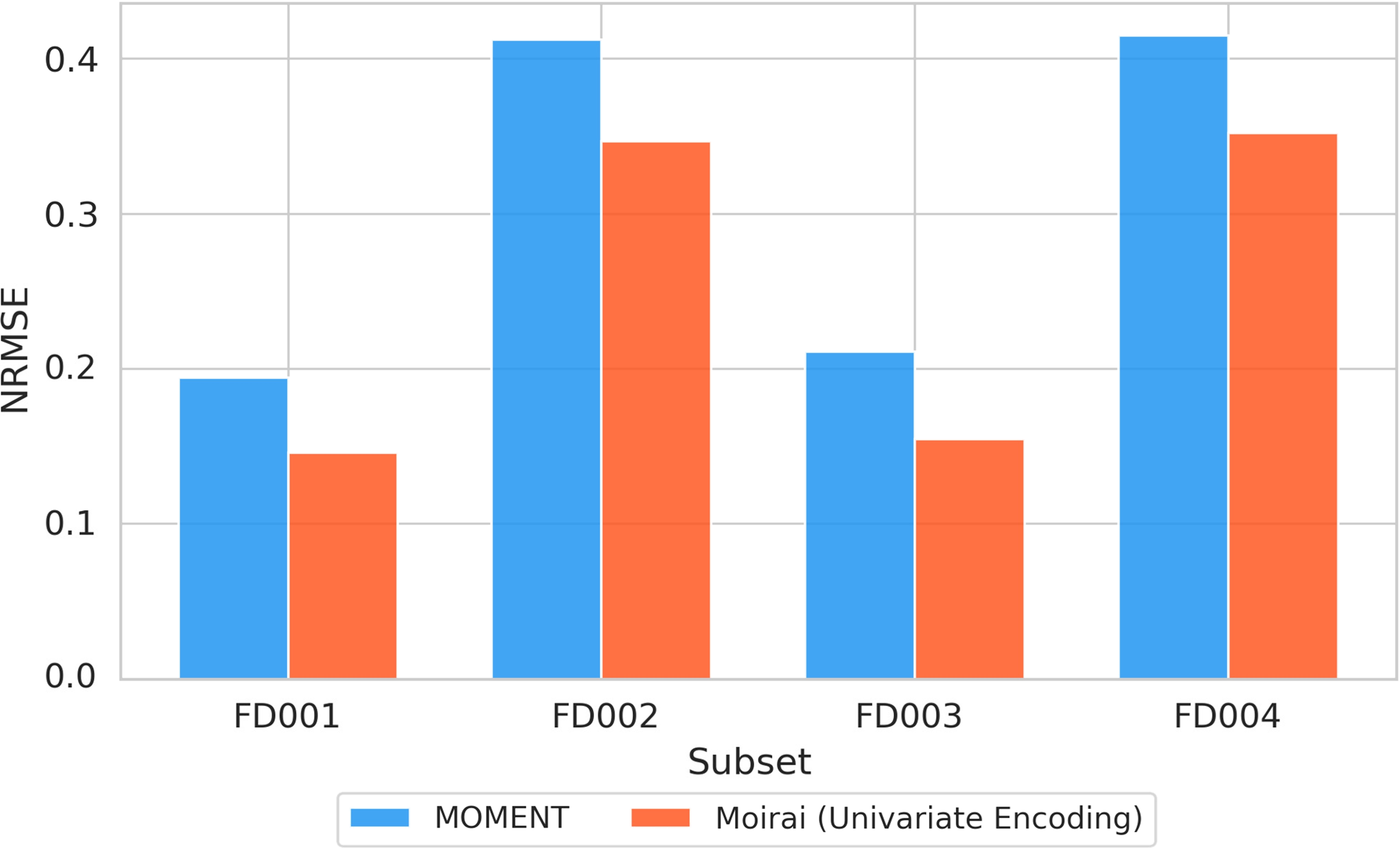}
    \centering
    \caption{Baseline NRMSE comparison between MOMENT and Moirai in univariate encoding mode across C-MAPSS subsets.}
    \label{fig:5}
\end{figure}

\textbf{Baseline performance highlights backbone differences.} Here we characterize the NRMSE of each backbone when all three cross-channel dependency sources are disabled, corresponding to the bottom rows of the MOMENT and Moirai blocks in Table~\ref{tab:7}. Figure~\ref{fig:5} shows the resulting NRMSE per C-MAPSS subset. Before any source of cross-channel dependencies is introduced, the backbone alone already accounts for a part of downstream performance. Moirai reaches a weighted average NRMSE of 0.2930 against MOMENT's 0.3444, and is better on all four subsets by margins well outside the fold-to-fold spread. The difference grows with operational complexity: it averages 0.064 NRMSE on the multi-condition subsets against 0.019 on the single-condition ones, or 16\% against 11\% in relative terms. In this configuration every source of cross-channel information is disabled by design, so neither backbone yet has access to the sensor relationships that separate degradation from operating-regime variation.

\begin{figure}[t!]
    \includegraphics[width=.6\linewidth]{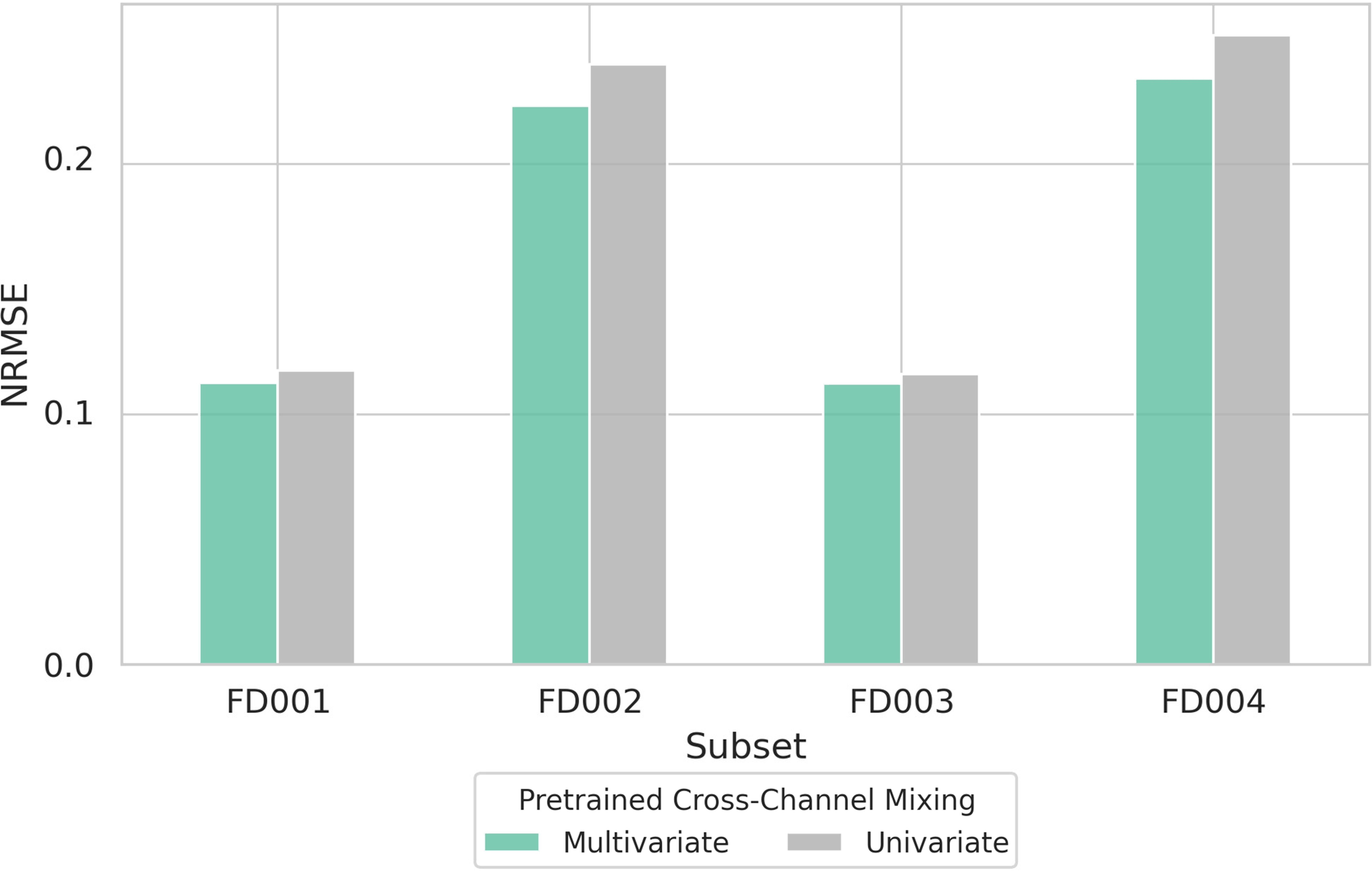}
    \centering
    \caption{Marginal effect of cross-channel dependencies for Moirai learnt from large-scale pretraining.}
    \label{fig:6}
\end{figure}

\textbf{Pretrained attention provides consistent gains, and larger ones where operating conditions vary.} Here we isolate the marginal contribution of pretrained cross-channel mixing by comparing Moirai rows with pretrained cross-channel mixing enabled against those with it disabled in Table~\ref{tab:7}, averaged across all target-task adaptation and digital twin fusion settings. Figure~\ref{fig:6} shows mean NRMSE under each condition per C-MAPSS subset. For Moirai, enabling the cross-channel dependencies modeled by pretrained weights allows all 14 sensor channels to interact through the model's Any-Variate attention mechanism during the forward pass, leveraging cross-channel dependency learned during pre-training on diverse time-series corpora. This reduces NRMSE in all six configuration pairs on every subset. The reduction is about 7\% on the multi-condition subsets and about 4\% on the single-condition ones. Under a single operating regime, each sensor channel already tracks degradation in a stable way, so encoding the channels independently loses little. When six regimes are interleaved, the operating point is expressed in the pattern across channels rather than in any one of them, and joint encoding gives the model access to information that channel-independent encoding cannot represent. This indicates that the cross-channel structure carried in the pretrained weights is most useful where degradation has to be separated from operational variation.

\begin{figure*}[b!]
    \centering
    \begin{subfigure}{\linewidth}
        \centering
        \includegraphics[width=.6\linewidth]{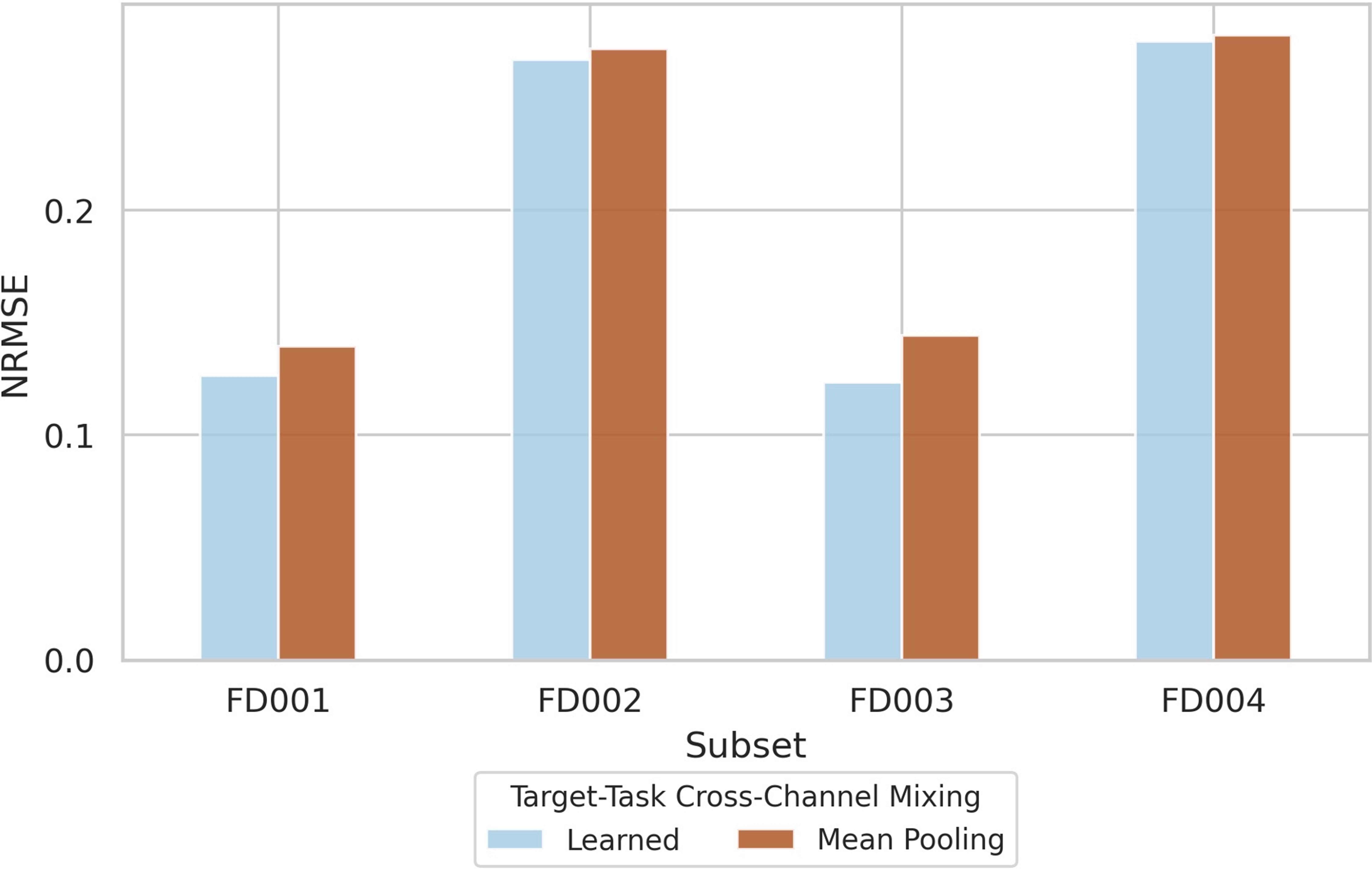}
        \caption{}
        \label{fig:7a}
    \end{subfigure}

    \vspace{1em}

    \begin{subfigure}{\linewidth}
        \centering
        \includegraphics[width=.6\linewidth]{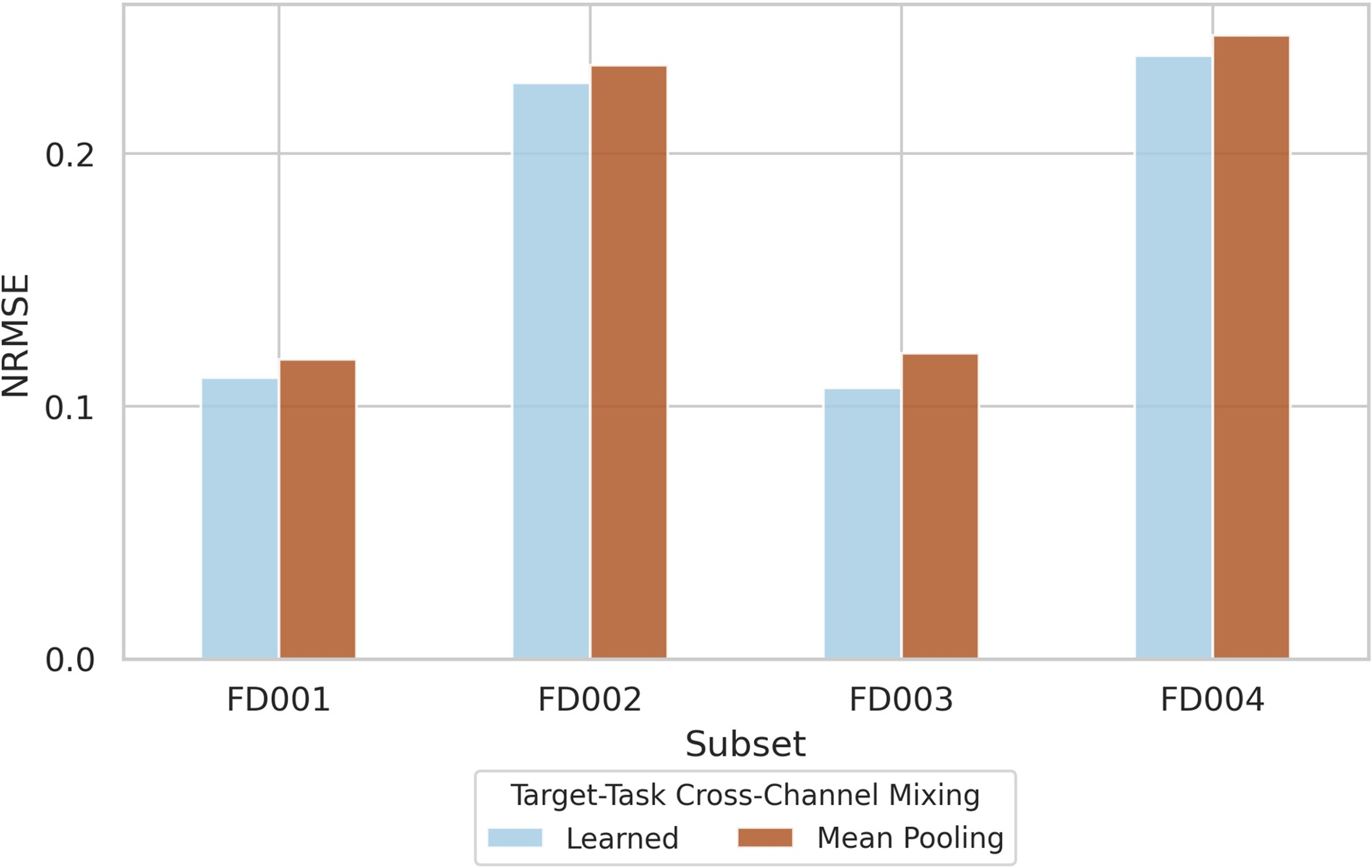}
        \caption{}
        \label{fig:7b}
    \end{subfigure}

    \caption{Marginal effect of target-task cross-channel dependencies for (a) MOMENT and (b) Moirai.}
    \label{fig:7}
\end{figure*}

\textbf{Learned target covariance contributes most under a single operating regime.} Here we isolate the marginal contribution of target-task channel mixing by comparing rows with learned channel weights against rows with mean pooling for both MOMENT and Moirai in Table~\ref{tab:7}, averaged across all digital twin fusion settings. Figure~\ref{fig:7} shows mean NRMSE under each condition per C-MAPSS subset for each backbone. Enabling cross-channel dependencies learned through target-task adaptation allows the model to learn a task-specific channel-mixing layer from the C-MAPSS training data via a depthwise convolution layer applied across sensor embeddings. This reduces NRMSE by 2.9\% for MOMENT and 4.7\% for Moirai when averaged across all digital twin fusion scenarios, and the reduction is concentrated on the single-condition subsets. For MOMENT it averages 6.6\% on the single-condition subsets against 1.4\% on the multi-condition ones, and for Moirai 8.8\% against 3.1\%. The depthwise convolution learns one set of channel-mixing weights per subset and applies them to every window regardless of the operating point. Under a single regime the sensors co-vary in the same way throughout the data, so one set of weights describes that relationship well. Under six regimes the same degradation produces different relative sensor deviations at different operating points, and one set of weights can only represent an average over the six, which describes none of them.

\begin{figure*}[t!]
    \centering
    \begin{subfigure}{\linewidth}
        \centering
        \includegraphics[width=.6\linewidth]{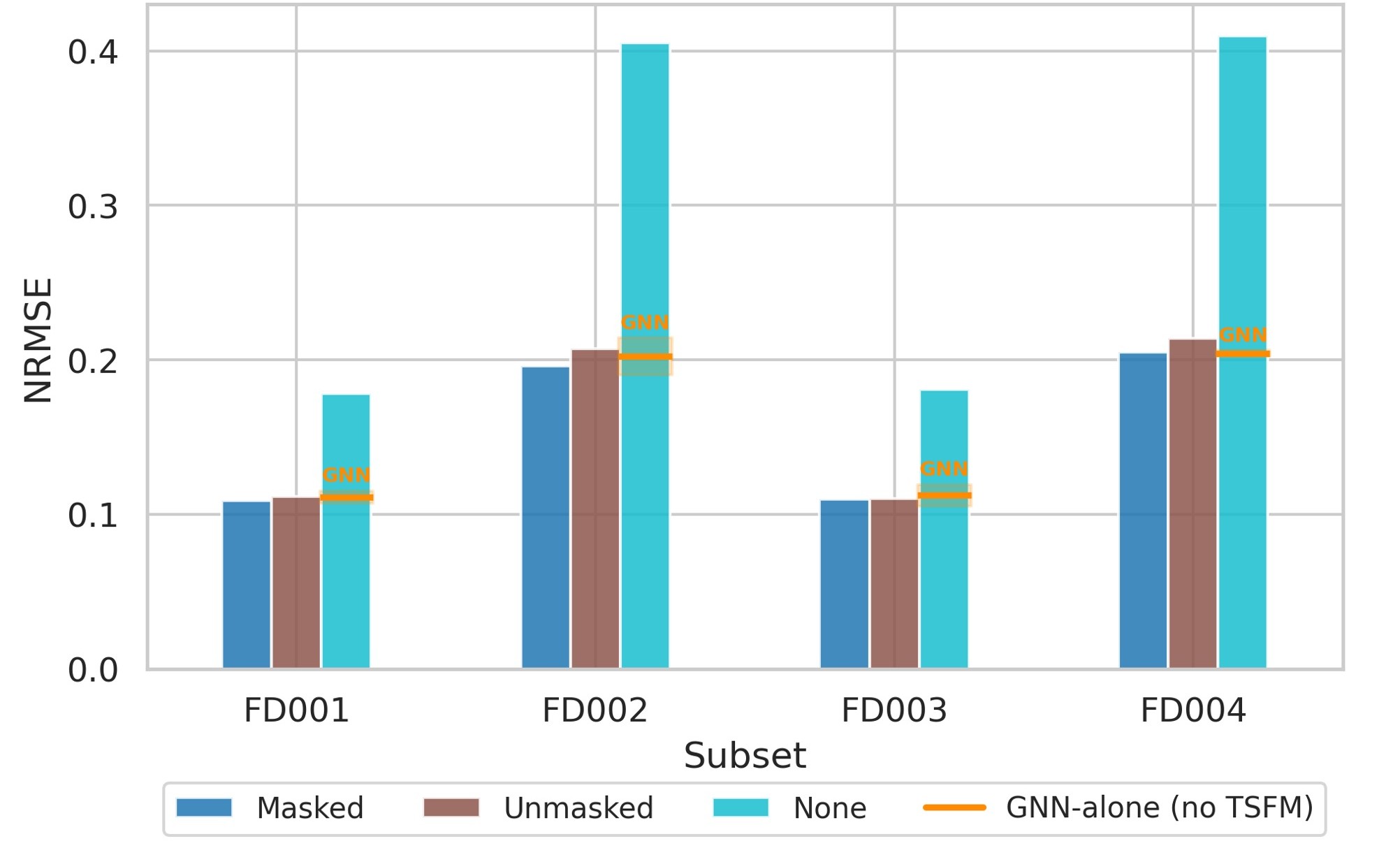}
        \caption{}
        \label{fig:8a}
    \end{subfigure}

    \vspace{1em}

    \begin{subfigure}{\linewidth}
        \centering
        \includegraphics[width=.6\linewidth]{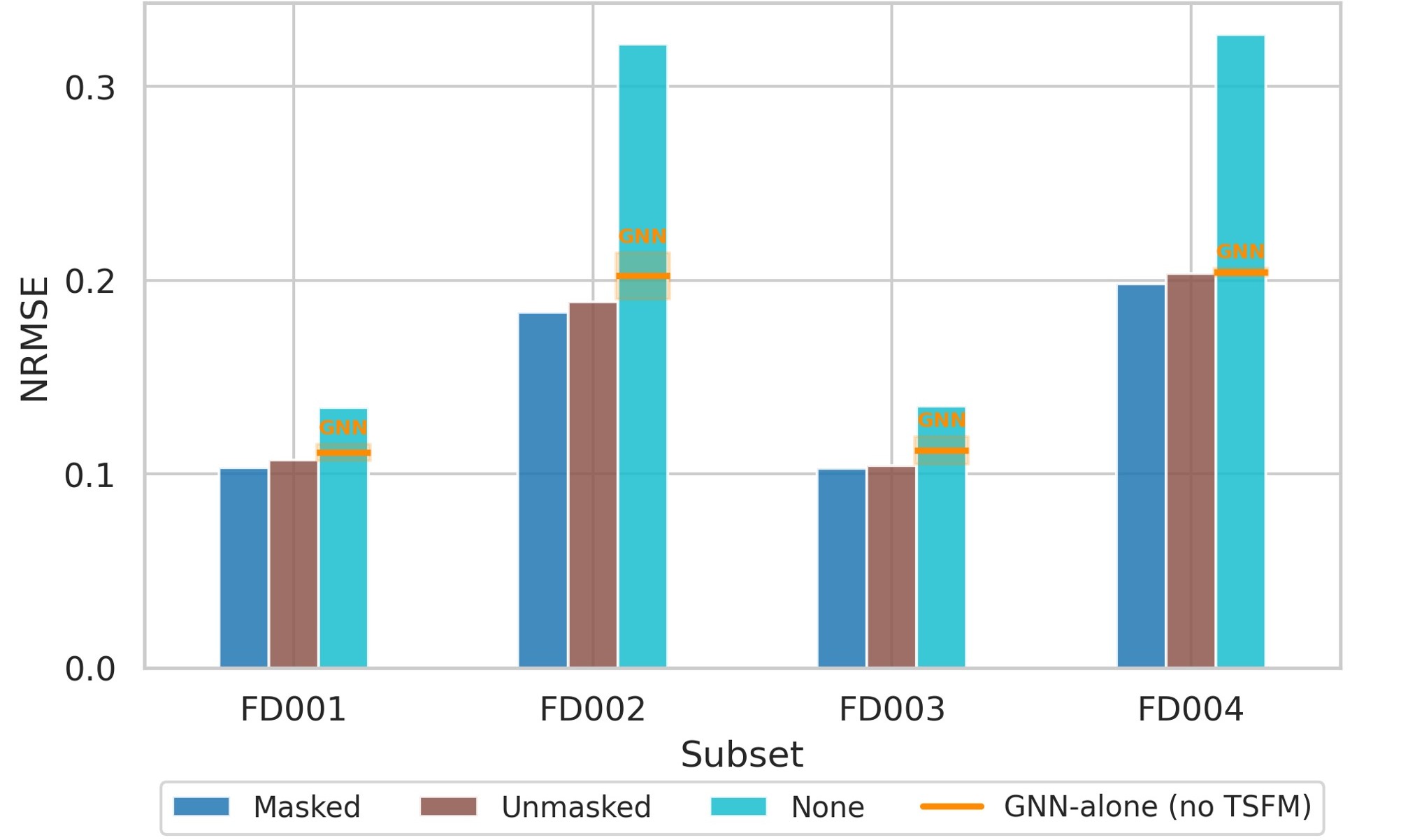}
        \caption{}
        \label{fig:8b}
    \end{subfigure}

    \caption{Marginal effect of digital twin fusion for (a) MOMENT and (b) Moirai.}
    \label{fig:8}
\end{figure*}

\textbf{Masked digital twin fusion outperforms unconstrained fusion and either source alone.} Here we isolate the marginal contribution of DT-derived topology by comparing masked fusion, unmasked fusion, and no fusion rows in Table~\ref{tab:7}, averaged across all Pretrain and Target-Task settings, using the GNN-alone row as reference. Figure~\ref{fig:8} shows mean NRMSE under each condition per C-MAPSS subset for each backbone. Integrating the component embeddings that the hierarchical GNN derives from the sensor-component couplings into the TSFM representation via cross-attention yields improvements across all scenarios, though much of the improvement over the TSFM alone reflects the predictive performance of the GNN itself rather than the connectivity it encodes. The orange reference lines in Figure~\ref{fig:8} indicate GNN-alone performance, which relies solely on message passing over the topological graph without any temporal features. Masked digital twin fusion, which constrains sensor embeddings to attend only to their physically associated components and immediate topological neighbors, reduces NRMSE by 6.7\% for Moirai and 1.5\% for MOMENT relative to that reference. Both backbones are ahead of it on the single-condition and the multi-condition subsets alike, Moirai by 7.8\% and 6.1\% and MOMENT by 2.0\% and 1.3\%, which confirms that the combination uses information from both sources rather than selecting the stronger one. The contribution of the mask itself, measured against unmasked fusion over the identical architecture, is 4.2\% for MOMENT and 2.7\% for Moirai in weighted average NRMSE, a limited effect consistent in direction with the connectivity comparison of Table~\ref{tab:6}. Constraining attention to physical connectivity prevents the model from using sensor-component correlations that hold under one operating regime and not another, and there are more such correlations to exclude when six regimes are interleaved. Unmasked fusion, by contrast, falls below the GNN-alone reference for MOMENT while remaining above it for Moirai: Moirai's pretrained attention already weights which channel interactions are informative, whereas MOMENT has no such mechanism, and the topological mask supplies that selection in its place.
 
\begin{figure}[t!]
    \includegraphics[width=.6\linewidth]{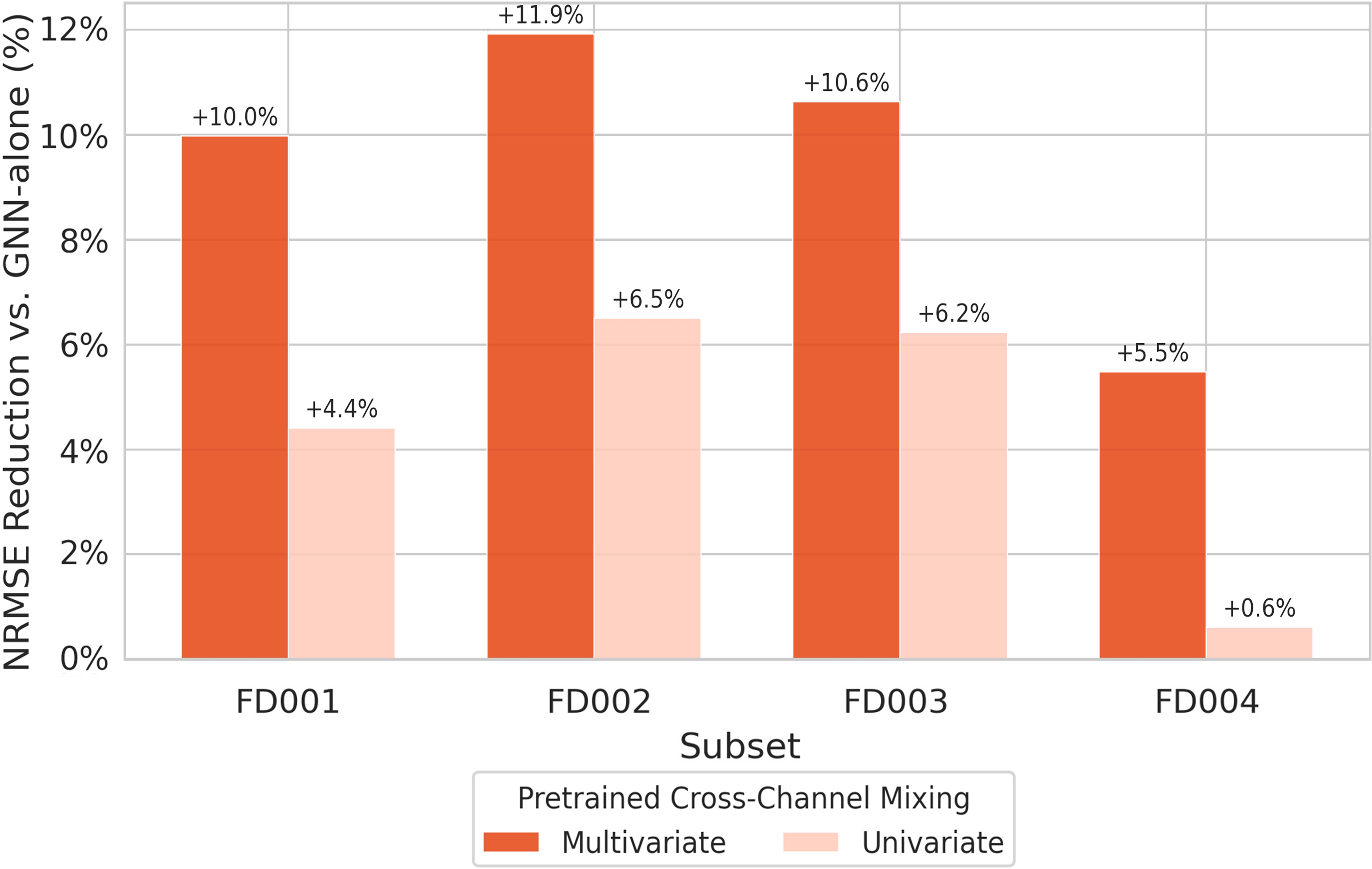}
    \centering
    \caption{NRMSE reduction relative to GNN-alone (baseline) under masked digital twin fusion.}
    \label{fig:9}
\end{figure}

\begin{figure*}[t!]
    \centering
    \begin{subfigure}{\linewidth}
        \centering
        \includegraphics[width=.6\linewidth]{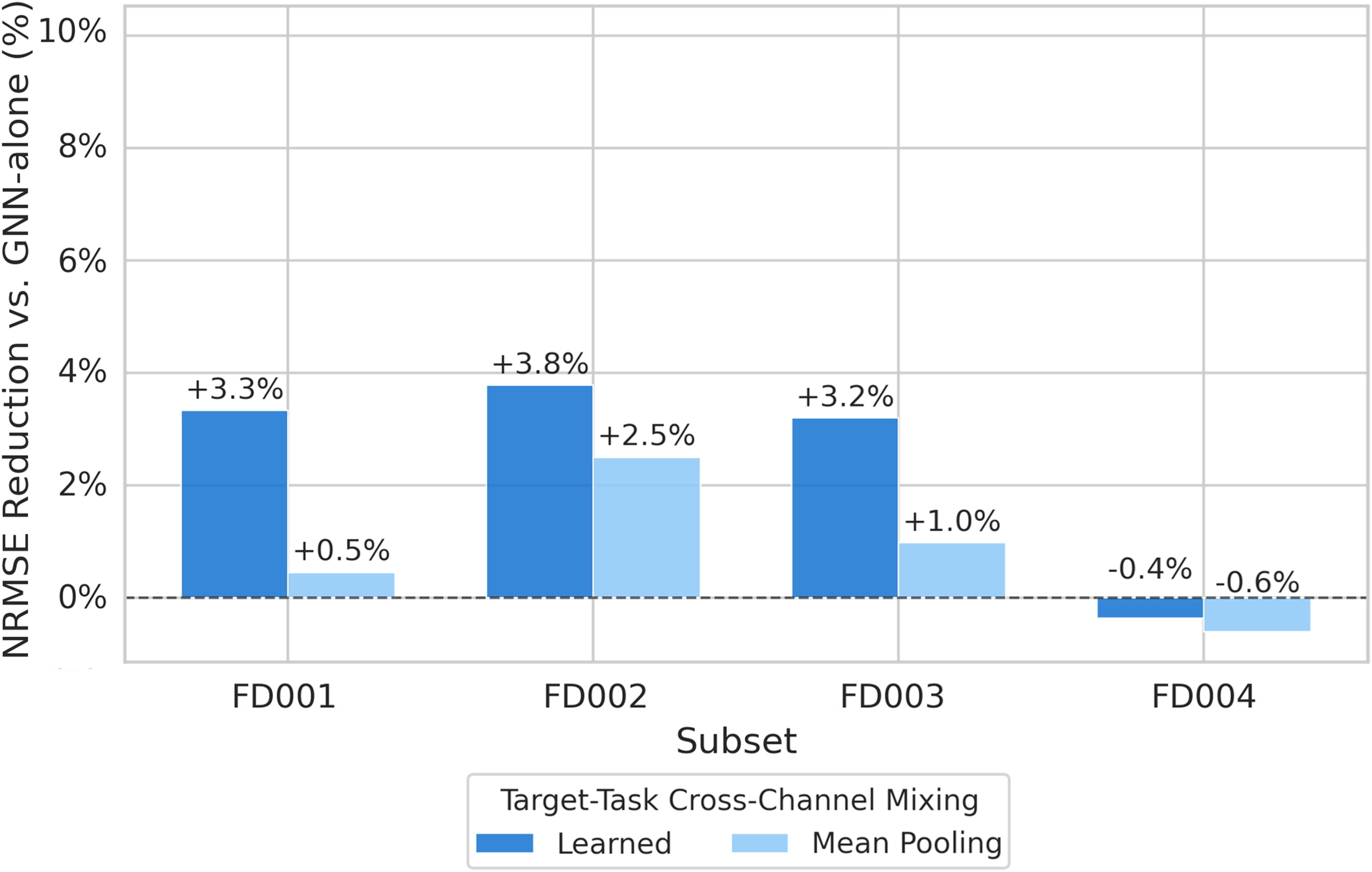}
        \caption{}
        \label{fig:10a}
    \end{subfigure}

    \vspace{1em}

    \begin{subfigure}{\linewidth}
        \centering
        \includegraphics[width=.6\linewidth]{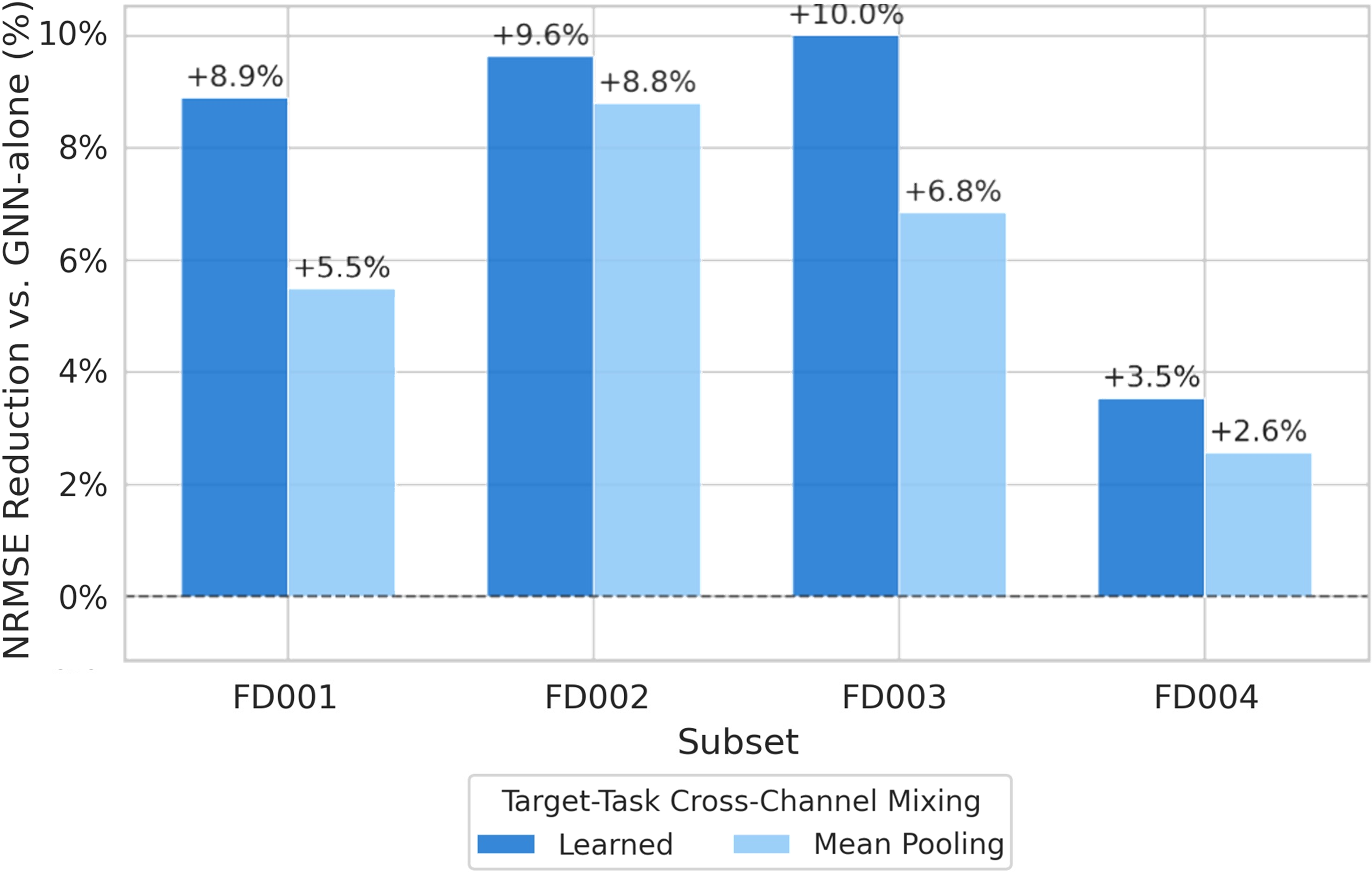}
        \caption{}
        \label{fig:10b}
    \end{subfigure}

    \caption{NRMSE reduction relative to GNN-alone (baseline) under masked digital twin fusion for (a) MOMENT and (b) Moirai.}
    \label{fig:10}
\end{figure*}

\textbf{Digital twin representation is complementary towards both cross-channel dependencies learned through pretrained weights and target-task adaptation.} Here we examine how DT-derived fusion interacts with each of the other two sources by measuring the additional NRMSE reduction over GNN-alone under masked fusion when pretrained cross-channel mixing or target-task adaptation is enabled versus disabled. Figures~\ref{fig:9} and~\ref{fig:10} show these interaction effects per C-MAPSS subset. For Moirai, masked digital twin fusion improves on the GNN-alone reference by 9.2\% when pretrained cross-channel mixing is enabled and by 4.1\% when it is disabled, a difference of 5.1 percentage points that is close to uniform across the benchmark, falling between 4.4 and 5.6 points on every subset. The topology therefore contributes the same amount whether or not the backbone brings its own account of how the channels relate, and the two add rather than substitute. The interaction with target-task adaptation is smaller and uneven. Enabling learned channel weights raises the gain from masked fusion by 1.3 percentage points for MOMENT and 1.6 for Moirai on a weighted average, and the increase averages 2.6 and 3.4 points on the single-condition subsets against 0.8 and 0.9 on the multi-condition ones. Target-task adaptation contributes little on the multi-condition subsets to begin with, so there is little for it to add on top of digital twin fusion there. Both interactions are positive on all four subsets and for both backbones, so the digital twin representation adds to what the other two sources provide rather than duplicating it.

\textbf{Proposed method reaches the best reported RMSE on single-condition subsets and closes most of the remaining gap where operating conditions vary.} Table~\ref{tab:8} shows the performance of our proposed model against state-of-the-art architectures. On the single-condition subsets the proposed method reaches the lowest RMSE reported. Every value reported here comes from ten models trained on different nine of the ten folds, whereas the cited results are obtained by training on the full training set of each subset, so the comparison does not favor the proposed method. On the multi-condition subsets it closes the distance between the off-the-shelf backbone benchmarked in Section~\ref{sec:4} and the state-of-the-art baselines, without reaching it. Several of those methods have access to the operating condition itself, either as an explicit input variable or implicitly through condition-specific normalization, while the models here see only the 14 sensor measurements. The asset structure records which sensors are attached to which components, and that relationship holds regardless of the operating point, but it carries no information about which regime produced a given reading. The digital twin representation therefore supplies health-related information that is complementary to what the frozen backbone provides, and under a single operating regime it is sufficient to match or exceed models trained end-to-end for this task; where six regimes are interleaved it removes most of the remaining difference but not all of it.

\begin{table}[t!]
\small
\centering
\caption{Comparison of proposed method against state-of-the-art RUL prediction baselines. For the proposed method, the value reported is the mean over the ten folds, under the protocol stated in Section~\ref{sec:4.1}; baseline values are cited from existing works. Bold marks the best value in each column; the proposed method is shown in the shaded row.}
\label{tab:8}
\begin{tabular}{lcccccccc}
\hline
 & \multicolumn{2}{c}{\textbf{FD001}} & \multicolumn{2}{c}{\textbf{FD002}} & \multicolumn{2}{c}{\textbf{FD003}} & \multicolumn{2}{c}{\textbf{FD004}} \\
\textbf{Method} & RMSE & Score & RMSE & Score & RMSE & Score & RMSE & Score \\
\hline
SVR~\cite{sateesh2016deep}             & 20.96 & 1382 & 42.00 & 589900 & 21.05 & 1598 & 45.35 & 37114 \\
CNN~\cite{sateesh2016deep}             & 18.45 & 1287 & 30.29 & 13570  & 19.82 & 1596 & 29.16 & 7886  \\
LSTM~\cite{zheng2017long}              & 16.14 & 338  & 24.49 & 4450   & 16.18 & 852  & 28.17 & 5550  \\
AGCNN~\cite{liu2021life}               & 12.42 & 226  & 19.43 & 1492   & 13.39 & 227  & 21.50 & 3392  \\
MCLSTM~\cite{xiang2021deep}            & 13.71 & 315  & ---   & ---    & ---   & ---  & 23.81 & 4826  \\
BiGRU-TSAM~\cite{zhang2022prediction}  & 12.56 & 213  & 18.94 & 2264   & 12.45 & 233  & 20.47 & 3610  \\
DRLRULe~\cite{hu2023remaining}         & 12.17 & \textbf{208}  & \textbf{16.28} & \textbf{1437}   & 13.09 & \textbf{226}  & \textbf{18.87} & \textbf{1726}  \\
\hline
\rowcolor{gray!15}
Proposed (Moirai) & \textbf{11.58} & 215 & 21.15 & 4619 & \textbf{11.77} & 272 & 22.89 & 5301 \\
\hline
\end{tabular}%
\end{table}

\section{Conclusion} \label{sec:7}
This work investigated how much cross-channel information each of three sources, pretrained weights, target-task adaptation, and digital twin-derived topology, contributes to general-purpose TSFMs, and whether these contributions are complementary. A benchmark of five frozen TSFMs on C-MAPSS RUL prediction showed that multivariate architectures substantially outperform univariate ones, particularly under varying operating conditions, while dedicated end-to-end models remain ahead of every off-the-shelf backbone. We then proposed a topology-informed fusion approach in which the asset structure a digital twin records constrains cross-attention between TSFM embeddings and component representations, and an ablation study across the three sources showed that they are complementary rather than redundant, with topology-constrained attention outperforming unconstrained fusion by a limited margin. A test of the connectivity itself showed that replacing the ground truth sensor-component structure with alternatives of the same size degrades performance, more so where operating conditions vary; the structure a digital twin records therefore carries health-related information, though its contribution in this setting is limited. A frozen multivariate TSFM informed by this structure matches or exceeds dedicated end-to-end models on the single-condition subsets and closes roughly four-fifths of the remaining distance on the multi-condition ones, on a regression task fundamentally different from its pretraining objectives, providing initial empirical grounding for the position that digital twin representations enhance the generalizability of TSFMs on unseen downstream tasks.

These findings rest on a single task and a single asset type, drawn from one dataset and a limited set of backbones; whether the approach yields comparable gains on other tasks and in other industrial domains, and how robust the effect is in general, remains to be established. The connectivity comparison also does not isolate the topology completely, since the GNN is trained on the sensor readings and its performance reflects more than the injected structure, and the graph itself remains static and binary \cite{ying2019hierarchical}. The topology is, moreover, only one of the information and simulation models a digital twin maintains; how much each adds, and how each should be encoded to inform a frozen backbone, is the question our ongoing work takes up. Extending TSFMs to interpret analytical objectives expressed in natural language would further move digital twin platforms from executing predefined tasks toward answering varied operational queries \cite{ma2025autonomy}.

\section*{Data Availability Statement}
The data that support the findings of this study are available in the NASA Open Data Portal at \url{https://data.nasa.gov/dataset/cmapss-jet-engine-simulated-data}. These data were derived from the following resources available in the public domain: CMAPSS Jet Engine Simulated Data (CMAPSSData.zip), \url{https://data.nasa.gov/docs/legacy/CMAPSSData.zip}.

\section*{CRediT Author Contributions}
Sizhe Ma: Conceptualization, Data curation, Formal analysis, Investigation, Methodology, Software, Validation, Visualization, Writing - Original draft, Writing - Review and editing. Katherine A. Flanigan: Conceptualization, Funding Acquisition, Methodology, Project administration, Resources, Supervision, Writing - Review and editing. Mario Berg\'es: Conceptualization, Funding Acquisition, Methodology, Project administration, Resources, Supervision, Writing - Review and editing.

\section*{Acknowledgements}
This material is based upon work supported by the National Science Foundation under Grant No. 2527269 and the Federal Railroad Administration (FRA), U.S. Department of Transportation under Contract \#693JJ623C000020.

\appendix
\setcounter{table}{0}
\setcounter{figure}{0}
\section{Pretraining Corpora of the Evaluated Backbones} \label{app:1}
 
Table~\mbox{\ref{tab:a1}} shows the pretraining corpus of each backbone evaluated in Section~\mbox{\ref{sec:4}} and Section~\mbox{\ref{sec:6}}, as described by its source publication, together with how far that corpus goes toward familiarity with the data used in this study. We grade that familiarity using the three categories introduced in existing work assessing the readiness of TSFMs for a target application \mbox{\cite{mulayim2024ready}}:
 
\begin{itemize}
\item \textit{Modality.} The corpus contains data of the same measurement modality as the test channels.
\item \textit{Dynamics.} The corpus contains time series generated by dynamical processes similar to those governing the test data.
\item \textit{Dataset.} The corpus contains the test dataset itself.
\end{itemize}
 
Instantiated for the present study, the modalities are the gas-path temperatures, pressures, rotational speeds and flow-derived quantities of Table~\mbox{\ref{tab:3}}; the governing dynamics are the thermodynamic operation of a turbofan engine and its progressive degradation toward functional failure; and the dataset is C-MAPSS.

All five corpora reach modality-level familiarity and no further. Each contains measurements of physical quantities that also appear among the C-MAPSS channels, principally temperature and pressure, so the backbones are not encountering these signals for the first time; the rotational-speed and bleed-flow channels particular to gas-path instrumentation are not part of any pretraining corpus. No corpus reaches dynamics-level familiarity, since none is documented as containing time series governed by the operation or degradation of thermo-mechanical machinery, and none reaches dataset-level familiarity, since none of them include C-MAPSS.
 
\begin{table}[t!]
\footnotesize
\centering
\begin{threeparttable}
\caption{Pretraining corpora of the five evaluated TSFMs, as reported by their source publications.}
\label{tab:a1}
\setlength{\tabcolsep}{5pt}
\renewcommand{\arraystretch}{1.2}
\begin{tabular}{@{}l >{\raggedright\arraybackslash}p{0.30\textwidth} >{\raggedright\arraybackslash}p{0.21\textwidth} c@{}}
\toprule
\textbf{Model} & \textbf{Pretraining corpus} & \textbf{Domains} & \textbf{\begin{tabular}{@{}c@{}}Familiarity\\ level\end{tabular}} \\
\midrule
MOMENT \cite{goswami2024moment} &
Time Series Pile: Informer, Monash, UCR/UEA, TSB-UAD &
Energy, transport, climate, healthcare, web &
Modality \\
Moirai \cite{woo2024unified} &
LOTSA: BuildingsBench, LargeST, ERA5, Monash, cluster traces &
Energy, transport, climate, cloud, web, retail &
Modality \\
Lag-Llama \cite{rasul2023lagllama} &
Monash, ETT, UCI, KDD Cup 2018, cluster traces &
Energy, transport, cloud, nature, economics &
Modality \\
Chronos \cite{ansari2024chronos} &
Monash, M-competitions, Kaggle; plus synthetic (TSMixup, KernelSynth) &
Energy, transport, climate, healthcare, retail, web &
Modality \\
TimesFM \cite{das2024decoder} &
Wikipedia, Google Trends, M4, Electricity, Traffic, Weather; plus synthetic &
Web, economics, energy, transport, climate &
Modality \\
\bottomrule
\end{tabular}
\end{threeparttable}
\end{table}

\section{Structures Drawn for the Ablation Study} \label{app:2}
\setcounter{table}{0}

Section~\ref{sec:6.1} compares the ground truth sensor-component structure against two alternatives. This appendix lists the structure drawn for each of the ten folds.
 
Any single structure representing inaccurate couplings among components and sensors might be wrong in an unrepresentative way, so rather than fixing on one, each fold draws at random from a fixed set of coupling structures. The first probes the couplings among components: ten sets are drawn at random from those that leave every component with as many couplings, and the graph with as many sensor connections, as the ground truth structure. Table~\ref{tab:b1} lists them. The second probes the sensor grounding: ten are drawn at random from the assignments that move every sensor to a different component and leave every component with as many sensors. Table~\ref{tab:b2} lists them.
 
\begin{table}[H]
\small\centering
\caption{Component couplings under the ground truth structure and under the ten alternatives, one per fold (F1--F10). A mark indicates the pair is coupled.}
\label{tab:b1}
\resizebox{\textwidth}{!}{%
\begin{tabular}{l|c|cccccccccc}
\hline
\textbf{Component pair} & \textbf{\begin{tabular}{@{}c@{}}Ground\\truth\end{tabular}} & \textbf{F1} & \textbf{F2} & \textbf{F3} & \textbf{F4} & \textbf{F5} & \textbf{F6} & \textbf{F7} & \textbf{F8} & \textbf{F9} & \textbf{F10} \\
\hline
Fan--LPC & \cmark &  &  &  &  &  &  &  & \cmark &  &  \\
Fan--HPC &  & \cmark & \cmark & \cmark &  &  & \cmark & \cmark &  & \cmark & \cmark \\
Fan--Comb. &  &  &  & \cmark &  &  &  &  &  &  &  \\
Fan--HPT &  & \cmark &  &  & \cmark &  & \cmark & \cmark &  & \cmark & \cmark \\
Fan--LPT & \cmark &  & \cmark &  & \cmark & \cmark &  &  & \cmark &  &  \\
Fan--Core &  &  &  &  &  & \cmark &  &  &  &  &  \\
LPC--HPC & \cmark &  &  & \cmark & \cmark & \cmark & \cmark & \cmark &  & \cmark & \cmark \\
LPC--Comb. &  &  & \cmark &  &  &  & \cmark & \cmark &  &  &  \\
LPC--HPT &  & \cmark & \cmark & \cmark & \cmark & \cmark & \cmark &  & \cmark &  & \cmark \\
LPC--LPT & \cmark & \cmark & \cmark &  &  &  &  & \cmark &  & \cmark & \cmark \\
LPC--Core &  & \cmark &  & \cmark & \cmark & \cmark &  &  & \cmark & \cmark &  \\
HPC--Comb. & \cmark & \cmark & \cmark &  & \cmark & \cmark &  &  & \cmark & \cmark & \cmark \\
HPC--HPT & \cmark & \cmark & \cmark & \cmark &  & \cmark & \cmark & \cmark & \cmark & \cmark &  \\
HPC--LPT &  & \cmark &  & \cmark & \cmark & \cmark &  &  & \cmark &  &  \\
HPC--Core & \cmark &  & \cmark &  & \cmark &  & \cmark & \cmark & \cmark &  & \cmark \\
Comb.--HPT & \cmark & \cmark &  &  & \cmark & \cmark &  &  & \cmark & \cmark &  \\
Comb.--LPT &  &  &  & \cmark &  &  & \cmark & \cmark &  &  & \cmark \\
HPT--LPT & \cmark &  & \cmark & \cmark & \cmark & \cmark & \cmark & \cmark & \cmark & \cmark & \cmark \\
HPT--Core & \cmark &  & \cmark & \cmark &  &  &  & \cmark &  &  & \cmark \\
LPT--Core &  & \cmark &  &  &  &  & \cmark &  &  & \cmark &  \\
\hline
Couplings retained & 10 & 4 & 7 & 4 & 6 & 6 & 4 & 6 & 7 & 6 & 6 \\
\hline
\end{tabular}
}
\end{table}
 
\begin{table}[H]
\small\centering
\caption{Sensor and component couplings under the ground truth structure and under the ten alternatives, one per fold (F1--F10).}
\label{tab:b2}
\resizebox{\textwidth}{!}{%
\begin{tabular}{l|c|cccccccccc}
\hline
\textbf{Sensor} & \textbf{\begin{tabular}{@{}c@{}}Ground\\truth\end{tabular}} & \textbf{F1} & \textbf{F2} & \textbf{F3} & \textbf{F4} & \textbf{F5} & \textbf{F6} & \textbf{F7} & \textbf{F8} & \textbf{F9} & \textbf{F10} \\
\hline
s2 & LPC & HPC & LPT & Fan & HPT & Fan & LPT & Core & Fan & Core & Core \\
s3 & HPC & Comb. & Comb. & LPC & LPC & LPC & LPC & Fan & LPT & Fan & LPT \\
s4 & LPT & Fan & HPT & Comb. & Fan & HPC & Core & Fan & HPC & LPC & HPC \\
s7 & HPC & Fan & Fan & Fan & Core & Fan & Fan & LPC & LPT & Fan & LPT \\
s8 & Fan & HPC & LPC & LPC & Comb. & Core & Core & HPC & HPC & LPC & HPC \\
s9 & Core & LPC & LPT & HPT & LPT & HPC & HPC & Fan & HPT & HPT & LPC \\
s11 & HPC & LPT & Fan & LPT & Fan & Fan & LPC & HPT & LPC & LPT & Fan \\
s12 & Comb. & LPC & Fan & HPC & LPT & LPT & Fan & LPT & Fan & HPC & LPC \\
s13 & Fan & Core & Core & HPC & Core & HPT & LPT & LPC & Core & Comb. & Core \\
s14 & Core & LPT & LPC & LPT & LPC & LPC & HPC & Comb. & HPC & Fan & HPT \\
s15 & Fan & Core & Core & Core & HPC & Comb. & Comb. & HPC & LPC & HPC & Comb. \\
s17 & LPC & HPT & HPC & Core & HPC & LPT & HPT & LPT & Comb. & LPT & Fan \\
s20 & HPT & HPC & HPC & Fan & Fan & HPC & Fan & HPC & Core & HPC & HPC \\
s21 & LPT & Fan & HPC & HPC & HPC & Core & HPC & Core & Fan & Core & Fan \\
\hline
Sensors unchanged & 14 & 0 & 0 & 0 & 0 & 0 & 0 & 0 & 0 & 0 & 0 \\
\hline
\end{tabular}
}
\end{table}

\bibliographystyle{abbrv}
\bibliography{reference}

\end{document}